\documentclass[useAMS,usenatbib,twocolumn]{mnras}
\usepackage{graphicx}
\usepackage{subfigure}
\usepackage[fleqn]{amsmath}
\usepackage{url}
\usepackage{soul}
\usepackage{xcolor}

\newcommand{\mpt}{\mathrm{.}}
\newcommand{\mcm}{\mathrm{,}}

\renewcommand{\Vec}[1]{ \mbox{\boldmath$ #1 $} }

\newcommand{\grale}{{\sc Grale}}
\newcommand{\lenstool}{{\sc Lenstool}}
\newcommand{\glafic}{{\sc glafic}}
\newcommand{\pixelens}{{\sc PixeLens}}
\newcommand{\swunited}{{\sc SWUnited}}
\newcommand{\lensperfect}{{\sc LensPerfect}}
\newcommand{\wslapplus}{{\sc WSLAP+}}
\newcommand{\sawlens}{{\sc SaWLens}}
\newcommand{\glee}{{\sc Glee}}
\newcommand{\WF}{1.0}

\newcommand{\WFTHREE}{0.333}

\title[Extended lens reconstructions with \grale]
{Extended lens reconstructions with \grale: exploiting time domain, substructural and weak-lensing information}
\author[J. Liesenborgs, L.L.R. Williams, J. Wagner and S. De Rijcke]
             {Jori Liesenborgs$^1$\thanks{Corresponding author: jori.liesenborgs@uhasselt.be},
              Liliya L.R. Williams$^2$,
              Jenny Wagner$^3$ and
              Sven De Rijcke$^4$\\
              $^1$ UHasselt -- tUL, Expertisecentrum voor Digitale Media, Wetenschapspark 2, B-3590, Diepenbeek, Belgium \\
              $^2$ School of Physics and Astronomy, University of Minnesota, 116 Church Street, Minneapolis, MN 55455, USA \\
              $^3$ Universit\"at Heidelberg, Zentrum f\"ur Astronomie, Astron. Rechen-Institut, M\"onchhofstr. 12–14, 69120 Heidelberg, Germany \\
              $^4$ Ghent University, Dept. Physics \& Astronomy, Krijgslaan 281, S9, B-9000, Ghent, Belgium \\
              }

\begin{document}

\date{} % TODO
	
\pagerange{\pageref{firstpage}--\pageref{lastpage}} \pubyear{2020}

\maketitle \label{firstpage} 
\begin{abstract} 
    The information about the mass density of galaxy clusters 
    provided by the gravitational lens effect has inspired
    many inversion techniques. In this article, updates to the previously
    introduced method in \grale{} are described, and explored in a number
    of examples. The first looks into a different way of incorporating
    time delay information, not requiring the unknown source position.
    It is found that this avoids a possible bias that leads
    to ``over-focusing'' the images, i.e. providing source position estimates 
    that lie in a considerably smaller region than the true positions.
    The second is inspired by previous reconstructions of the cluster of galaxies
    MACS~J1149.6+2223, where a multiply-imaged background galaxy contained a 
    supernova, SN Refsdal, of which four additional images were produced by 
    the presence of a smaller cluster galaxy. The inversion for the
    cluster as a whole, was not able to recover sufficient detail interior
    to this quad. We show how constraints on such different
    scales, from the entire cluster to a single member
    galaxy, can now be used, allowing such small scale substructures
    to be resolved.
    Finally, the addition of weak lensing information to this
    method is investigated. While this clearly helps 
    recover the environment around the strong lensing region, the mass sheet
    degeneracy may make a full strong and weak inversion difficult, depending
    on the quality of the ellipticity information at hand.
    We encounter ring-like structure at the boundary of the two
    regimes, argued to be the result of combining strong and weak lensing
    constraints, possibly affected by degeneracies.
\end{abstract}

\begin{keywords}
    gravitational~lensing:~strong -- gravitational~lensing:~weak -- methods:~data~analysis -- dark~matter --
    galaxies:~clusters:~general 
\end{keywords}

\section{Introduction}

    Due to their extended and particularly massive nature, gravitational 
    lensing by clusters of galaxies can provide various clues
    about their matter distributions. 
    In the so-called strong-lensing regime a massive central region
    can produce multiply imaged sources, currently exceeding 100 images in some
    cases, as in the study of Abell 1689 \citep{2005ApJ...621...53B},
    and many of the Hubble Frontier Fields clusters (HFF; \citet{Lotz_2017}).
    As first recognized by \citet{1990ApJ...349L...1T}
    in the context of galaxy lensing, further away from the centre there is
    still a systematic distortion in the shape of background galaxies,
    an effect described as weak lensing.
    Furthermore, the gravitational deflection of light also has an effect on
    the distribution of these background objects on the sky.

    While calculating the effect a known gravitational lens has on one or more 
    background objects is relatively straightforward, in practice one has only
    very little information about both sources and lens. The real life situation
    is therefore such that one has only observed the gravitational lensing
    effect, but wants to obtain information about the lensing mass distribution 
    as well as about the background sources, information that lies encoded in the observation. 
    The effects above depend on the precise distribution of the matter, both
    luminous and dark, and have in turn led to many so-called lens inversion methods
    attempting to reconstruct this distribution,
    varying in the kind of information they use as constraints, underlying
    assumptions about the mass model, and optimization techniques. 

    The methods using information from the strong, weak or both lensing regimes,
    can be classified as parametric, or non-parametric. The former, sometimes
    also referred to as simply parameterized models, pre-suppose a particular shape
    of the mass distribution, of which a relatively small number of parameters
    still needs to be optimized to match the observations; 
    examples include \lenstool{} \citep{2007NJPh....9..447J,2009MNRAS.395.1319J},
    \glafic{} \citep{2010PASJ...62.1017O}, \glee{} \citep{2010A&A...524A..94S,2015ApJ...800...38G} 
    and the light-traces-mass technique
    by \citet{2009ApJ...703L.132Z}. The other class, also called free-form methods,
    attempt to avoid a bias towards a particular shape of the mass density, typically
    needing a large number of parameters to do so. These can describe the mass density
    directly, like in \pixelens{} \citep{1997MNRAS.292..148S,2004AJ....128.2631W}, 
	\wslapplus{} \citep{2005MNRAS.362.1247D,2007MNRAS.375..958D,2014MNRAS.437.2642S}, 
    and the work of \citet{1998MNRAS.299..895B}, or alternatively indirectly using
    the so-called lensing potential. The weak-lensing only work of 
    \citet{1996ApJ...464L.115B} that parameterized this potential was extended to
    include strong lensing information through available image systems in
    in \swunited{} \citep{Bradac,2009ApJ...706.1201B}, and through estimates
    of the critical lines in \sawlens{} \citep{2006A&A...458..349C,2009A&A...500..681M}.
    Free-form inversion methods that model neither the potential nor the mass
    density directly include the strong lensing method \lensperfect{} \citep{LensPerfect},
    which considers models that precisely reproduce the images by exploring
    curl-free interpolations of the deflection field, and weak
    lensing methods based on \citet{1993ApJ...404..441K} that investigate
    how the mass distribution can be obtained directly from the measured deformation
    field. Accurate determination of the ellipticities of background galaxies
    form the cornerstone of weak lens inversions, stimulating comparisons of
    different techniques in e.g. the Shear TEsting Programme (STEP, \citet{2006MNRAS.368.1323H})
    and GRavitational lEnsing Accuracy Testing (GREAT) challenges (e.g. \citet{2014ApJS..212....5M}).

    In this article we describe additions to the strong lensing, non-parametric inversion 
    algorithm that was first introduced in \citet{Liesenborgs}, and was
    later christened\footnote{The name `\grale{}' is merely a contraction 
    of {\sc gra}vitational {\sc le}ns.}
    \grale. 
    The code\footnote{\url{https://research.edm.uhasselt.be/jori/grale2}}
    does not only include the aforementioned inversion algorithm to reconstruct
    the lensing mass density from observations, but also
    includes a variety of tools to perform and analyze simulations, that
    are helpful in evaluating the performance of the inversion procedure.

    The inversion method uses a genetic algorithm as the underlying
    optimization procedure, a technique from the wider class of evolutionary
    algorithms (see e.g \citet{EibenBook}) which are all inspired by the 
    way natural evolution produces individuals that are increasingly adapted to their 
    environment.
    While more traditional optimization techniques, like Markov Chain Monte Carlo (MCMC),
    explore the parameter space through a sequence of points that are
    by some metric adjacent, genetic algorithms allow the parameter space
    to be searched in a non-local way as well, by combining or exchanging
    parameters from multiple trial solutions (in biology, this corresponds
    to new chromosomes having properties based on both parents). The kinds
    of problems a genetic algorithm may handle, are less restricted, allowing
    e.g. combinatorial problems or problems with discrete parameter spaces to
    be tackled as well, as long as one can identify which solution is the
    better one from two or more trial solutions. An additional advantage,
    which will also be revisited later, is the way multiple objectives can
    be handled: classically, these are combined into a single number which
    is then optimized, but this requires one to carefully choose the weights
    for each objective as they are combined. In a multi-objective genetic
    algorithm, no such weights need to be assigned however. The major downside
    of this added flexibility is the lack of mathematical and statistical
    rigor. Not only are analyses of genetic algorithms only available in
    the simplest of cases, there is no guarantee that the solutions produced
    by the technique will be related to some desired probability distribution,
    as is the case with MCMC. 
    
    After summarizing the relevant lensing formalism in section \ref{sec:formalism},
    we describe this inversion procedure in section \ref{sec:grale}. Over the
    years, several modifications and extensions have been described, and
    for this reason the current state of the algorithm is first reviewed. 
    Further generalizations and additions are detailed, which are subsequently
    explored in the article. The first of these, an improved time delay
    criterion, is the subject of section \ref{sec:timedelays}. In 
    section \ref{sec:substruct}, the problem with the presence of small
    scale substructures, as well as possible solutions, is investigated,
    and in section \ref{sec:weak}
    the inclusion of weak lensing data is explored. The article concludes 
    with a discussion in section \ref{sec:discussion}. Unless otherwise
    stated, a flat $\Lambda\textrm{CDM}$ cosmological model is used 
    throughout the text
    with $H_0 = 70\,\mathrm{km}\,\mathrm{s}^{-1}\,\mathrm{Mpc}^{-1}$
    and $\Omega_m = 0.3$.

\section{Gravitational lensing formalism}\label{sec:formalism}

    Gravitational lensing is commonly modelled as a two-dimensional,
    projected mass distribution in a single so-called lens plane,
    which instantaneously deflects light rays over an angle \Vec{\hat{\alpha}}.
    This deflection causes points \Vec{\beta} in the source plane
    to be transformed into image plane points \Vec{\theta} according
    to the lens equation or ray-trace equation (see e.g. \citet{SchneiderBook}):
    \begin{equation}
        \Vec{\beta} = \Vec{\theta} - \frac{D_{\rm ds}}{D_{\rm s}} \Vec{\hat{\alpha}}(\Vec{\theta}) \mpt
    \end{equation}

    Here, $D_{\rm s}$ and $D_{\rm ds}$ represent the angular diameter
    distances from observer to source and lens to source respectively.
    Often, with a single reference source plane in mind, the scaled
    deflection angle $\Vec{\alpha} = D_{\rm ds}/D_{\rm s} \Vec{\hat{\alpha}}$
    is used instead; it is however important to keep in mind that this
    implicitly refers to a specific source distance. The deflection angle
    originates from the projected potential $\psi$:
    \begin{equation}
     \Vec{\alpha}(\Vec{\theta}) = \Vec{\nabla} \psi(\Vec{\theta}) \mpt
     \label{eq:gradpsi}
    \end{equation}

    It is of course the two-dimensional mass distribution $\Sigma(\Vec{\theta})$ that
    determines the deflection angle, and it can be shown that
    \begin{equation}
     \kappa(\Vec{\theta}) \equiv \frac{\Sigma(\Vec{\theta})}{\Sigma_{\rm crit}} = \frac{1}{2}\left(\frac{\partial \alpha_x}{\partial \theta_x} + \frac{\partial \alpha_y}{\partial \theta_y}\right) \mcm
    \end{equation}
    in which
    \begin{equation}
     \Sigma_{\rm crit} = \frac{c^2}{4\pi G D_{\rm d}} \frac{D_{\rm s}}{D_{\rm ds}} \mpt
    \end{equation} 
    In the equation above, $\kappa$ is called the convergence and $\Sigma_{\rm crit}$ 
    the critical density, both with respect to the source distance
    under consideration. Apart from the speed of light $c$ and the gravitational
    constant $G$, this last expression also contains $D_{\rm d}$, the angular
    diameter distance to the lens plane. 

    In the strong lensing regime, the lens equation above transforms a single
    source into multiple images. If the source itself is variable, these variations
    will appear at different times in different images. The time delay between
    image points $\Vec{\theta}_i$ and $\Vec{\theta}_j$ of the same source point
    $\Vec{\beta}$, is then given by $\Delta t_{ij} \equiv t(\Vec{\theta}_i, \Vec{\beta}) - t(\Vec{\theta}_j, \Vec{\beta})$
    where
    \begin{equation}
        t(\Vec{\theta},\Vec{\beta}) = \frac{1+z_d}{c} \frac{D_{\rm d} D_{\rm s}}{D_{\rm ds}} \left(\frac{1}{2}(\Vec{\theta}-\Vec{\beta})^2 - \psi(\Vec{\theta}) \right)
    \label{eq:timedelay}
    \end{equation}
    \citep{1985A&A...143..413S,SchneiderBook}.
    Here, the pre-factor also includes the redshift $z_d$ of the gravitational
    lens.

    From a first order approximation of the lens equation one obtains
    \begin{equation}
        \Delta\Vec{\beta} = \mathcal{A}(\Vec{\theta}) \cdot \Delta\Vec{\theta}\mcm
       \label{eq:firstorder}
    \end{equation}
    in which $\mathcal{A}$ is called the magnification matrix with elements
    \begin{equation}
        \mathcal{A}_{ij}(\Vec{\theta}) = \frac{\partial \beta_i}{\partial \theta_j} \mpt
    \end{equation}
    This matrix is also written as
    \begin{equation}
     \mathcal{A}(\Vec{\theta}) = (1-\kappa)
       \left(\begin{array}{cc}
           1 & 0 \\
           0 & 1
       \end{array}\right) - 
       \left(\begin{array}{cc}
           \gamma_1 & \gamma_2 \\
           \gamma_2 & -\gamma_1
       \end{array}\right) \mcm 
    \end{equation}
    showing a uniform scaling as well as deformation by the shear components $\gamma_1$ 
    and $\gamma_2$, where
    \begin{equation}
     \gamma_1 \equiv \frac{1}{2}\left(\frac{\partial \alpha_x}{\partial \theta_x} - \frac{\partial \alpha_y}{\partial \theta_y}\right)
    \textrm{, and }
    \gamma_2 \equiv \frac{\partial \alpha_x}{\partial \theta_y} = \frac{\partial \alpha_y}{\partial \theta_x}  \mpt
    \end{equation}

    In the weak lensing regime, one no longer has multiple images of the same source,
    but the deformations of background galaxies are still well described by these first order
    approximations, thereby providing information about convergence and shear. 
    Unfortunately, the information comes in the form of a combination,
    the so-called reduced shear $g_1$ and $g_2$, where
    \begin{equation}
     g_i = \frac{\gamma_i}{1-\kappa} \mpt
    \end{equation}

    The effect of weak lensing on source shapes is investigated in e.g. \citet{1995A&A...294..411S} 
    and \citet{1997A&A...318..687S}, and they show that using a complex notation
    where $g = g_1 + i g_2$, a source ellipticity $\epsilon^{{\rm (s)}} = \epsilon^{{\rm (s)}}_1 + i \epsilon^{{\rm (s)}}_2$
    is transformed into an image ellipticity $\epsilon$ according to
    \begin{equation}
     \epsilon = \left\{\begin{array}{cl}
        \dfrac{\epsilon^{{\rm (s)}} + g}{1 + g^*\epsilon^{{\rm (s)}}} & \textrm{if } |g| \le 1 \mcm \\
        \dfrac{1+g\epsilon^{{\rm (s)}*}}{\epsilon^{{\rm (s)}*} + g^*} & \textrm{if } |g| > 1 \mpt
    \end{array}\right. 
    \label{eq:elltrans}
    \end{equation}
    For an elliptical source with axes $a$ and $b$, where $b < a$, rotated over an
    angle $\varphi$, the source ellipticity would be
    \begin{equation}
     \epsilon^{{\rm (s)}} = \frac{1-b/a}{1+b/a} e^{i 2\varphi} 
     \label{eq:shearorient}
    \end{equation}
    with an analogous expression for the image ellipticity $\epsilon$.
    For a more general expression in terms of the quadrupole moments of the shape,
    the reader is referred to the aforementioned references. As is shown there,
    assuming that the source ellipticities average out to zero, the averaged image
    ellipticities then provide an estimate for the reduced shear:
    \begin{equation}
     \langle\epsilon\rangle = \left\{\begin{array}{cl}
        g & \textrm{if } |g| \le 1 \mcm \\
        \;\vspace{-0.3cm}\\
        \dfrac{1}{g^*} & \textrm{if } |g| > 1 \mpt
     \end{array}\right. 
     \label{eq:ellpred}
    \end{equation}
    A common approach is therefore to obtain observations of many background galaxies,
    determine their ellipticity values $\epsilon$,
    and average these to obtain estimates of the reduced shear.

    While both strong lensing information, i.e. the positions of multiple images of
    one or more sources, and weak lensing information, i.e. measured ellipticities
    of background galaxies, encode aspects of the mass distribution of the
    gravitational lens, unfortunately in practice degeneracies remain: multiple
    mass distributions will be equally acceptable reconstructions, but may differ
    in non-trivial ways. The most well known degeneracy is undoubtedly the mass
    sheet degeneracy \citep{FalcoMassSheet} which in the context of strong lensing
    is also called the steepness degeneracy \citep{SahaSteepness}.
    It was soon found to be a special case of several classes of invariance
    transformations that leave the observables in multiple-image configurations 
    invariant \citep{Gorenstein,2014A&A...564A.103S}. Making the lens reconstruction 
    independent of specific (parametric) lens models, it was found that these 
    degeneracies that had been treated as global transformations of the entire 
    lens and source plane properties, can be further generalised, such that they 
    locally apply to each system of multiple images individually 
    \citep{Liesenborgs3,2012MNRAS.425.1772L,2018A&A...620A..86W}. It is clear
    that these degeneracies cause difficulties in constraining
    the mass density of a specific lensing object at a certain redshift from 
    such lensing observations, with the integration of mass along the line of
    sight further confounding the issue \citep{2019MNRAS.487.4492W}.
    
    The core insight to understand the mass sheet degeneracy is that for a single 
    source distance, both a convergence $\kappa_0$ as well as the derived
    \begin{equation}
        \kappa_1(\Vec{\theta}) = \lambda\kappa_0(\Vec{\theta}) + (1-\lambda)
        \label{eq:msd}
    \end{equation}
    are compatible with the same image positions, and this for any choice of $\lambda$.
    The scale of the source plane is different however, where a larger mass sheet
    or less steep profile corresponds to a smaller source, scaled by a factor
    $\lambda$ in each dimension. 
    A similar relation for the lensing potentials causes the
    time delays between images to be rescaled as well, providing an opportunity
    to break this degeneracy if time delay measurements are available.

    The convergence refers to a specific source distance, and the simple construction
    above is therefore no longer available when multiple sources at different
    distances are involved. The more local variants of the degeneracy mentioned
	above however still allow similar degeneracies, only causing a difference
	in densities at the locations of the images, where a similar relation as
    equation (\ref{eq:msd}) still holds. Allowing minor deviations in the
    source to images correspondences only make this degeneracy even more difficult to break.

	The mass sheet degeneracy is of course not only a nuisance in strong lensing,
	but also in weak lensing as first described by \citet{1995A&A...294..411S}.
    If a single input shear field is used, a similar degeneracy
    as in (\ref{eq:msd}) is present, where the constant depends on the redshift
    distribution of the observed background ellipticities \citep{1997A&A...318..687S}.
    When the individual redshift information of these background sources is
    available however, \citet{2004A&A...424...13B} argue that the degeneracy can
    be lifted, at least in principle.

    The effect of such types of degeneracy is to rescale the source planes
    involved, and to modify the densities in a similar way as in (\ref{eq:msd}).
    For this reason, we will simply refer to this entire class of degeneracies
    as the mass sheet degeneracy (MSD), even when it is not restricted to a
    single source distance or corresponds to perfect rescalings.
    When comparing different gravitational lens models, the two effects,
    on source plane scales and densities, can be used to assess if the difference
    in models is due to the MSD. 
    The example model from Fig.~\ref{fig:tdreallenspt} that we shall encounter later
    illustrates this: the right hand panel of the reconstruction in the first
    row of Fig.~\ref{fig:tdreconstrallpt} shows a difference in source plane
    scale, where the corresponding top-left panel of Fig.~\ref{fig:tdrelativequant}
    shows how steepness and density offset, here sampled only at the positions 
    of the images, differ accordingly.

\section{Lens inversion with \grale}\label{sec:grale}

    \subsection{Genetic algorithm based inversion procedure}\label{sec:gainv}

        The inversion procedure that \grale{} uses, has been designed with strong
        lensing scenarios in mind. Being a non-parametric, or free-form inversion method, 
        there is no presupposed
        distribution of the lens plane mass; instead, the weights of a number of basis
        functions will be determined, thereby allowing a wide variety of projected 
        mass densities to be modelled. The amount, type and and location of the basis
        functions still need to be fixed, and to keep a handle on the complexities
        allowed in this regard, a strategy inspired by the work of \citet{2005MNRAS.362.1247D}
        was used. This approach starts by laying out basis functions on a uniform grid, and
        letting an optimization procedure determine their weights.
        Based on the resulting mass distribution a new grid is defined, in which regions
        with more mass are subdivided into finer grid squares. The optimization again tries
        to locate appropriate weights,
        and this entire procedure can be repeated a number of times as desired;
        usually after a number of subdivision steps the amount of weights becomes 
        larger than the observations can constrain, and the optimization procedure ceases
        to produce improvements. Figure 1 of \citet{Liesenborgs} illustrates this dynamic
        subdivision grid. In our approach, the basis functions used are projected Plummer
        spheres \citep{1911MNRAS..71..460P}, of which the width is set proportional to the width of a grid
        cell.

        A single inversion run thus consists of a number of steps that increase the
        resolution of the grid, where in each step an optimization procedure determines
        the weights of the basis functions. Inspired by the work of \citet{2005PASA...22..128B},
        a genetic algorithm (GA) is used as the optimization routine.
        As the name suggests, this optimization method mimics natural
        evolution, and starts by initializing a first set -- called a population --
        of random trial mass maps (randomly initialized weights of Plummer basis
        functions) -- often referred to as genomes or chromosomes.
        In a GA, each trial solution is assigned some measure for how successful
        the solution is -- called its fitness -- and a new population will be created
        by combining, cloning and mutating existing genomes. The key ingredient to evolve
        towards increasingly better solutions is to ensure that better trial solutions 
        create more offspring, i.e. to apply selection pressure. By default, the
        weights are all required to be positive, to ensure a positive mass density
        everywhere, although negative weights can be allowed as well, e.g. to provide
        corrections to a base model.

        In the original GA, a single fitness measure was used to estimate how
        compatible a trial mass distribution was with the observations, in essence by measuring the
        fractional overlap of the back-projected images. In a next step, it was found
        that the so-called null space could provide valuable information as well:
        the reconstruction should not only predict the observed images, but should
        also prevent the prediction of extra, unobserved images \citep{Liesenborgs2}.
        To handle two (or more) fitness criteria simultaneously, a multi-objective
        genetic algorithm (e.g. \citet{Deb}) is employed, thereby looking for a
        solution that optimizes several fitness criteria at the same time. In
        the most general multi-objective optimization there will be a trade-off between
        fitness measures. For example, if the first criterion would be how well the
        images can be predicted and a second criterion would encode how low the average
        density of the lens is, then a lens with zero density would optimize the
        second but clearly not the first; vice versa a mass distribution which is able
        to predict the images will certainly not have the lowest density. In our
        applications however, we employ fitness measures that are in this sense
        compatible with each other, that are believed to have an optimum at the same
        time. In the null space example there should exist a solution
        that not only predicts the observed images but also does not predict extra
        images that would have been detected in the observations.

        Within the GA, each trial solution or genome encodes the weights of the Plummer
        basis functions. More precisely, the stored values do not determine the Plummer
        weights directly, but only up to a certain scale factor. The genome therefore
        only describes the relative shape of the mass distribution. The scale factor
        to use for a genome is the one that produces the best overlap fitness
        of the back-projected images. In
        case a multi-objective GA is used and multiple fitness criteria are present,
        the scale is still fixed based on the overlap of back-projected images, and
        once this is obtained the other fitness values are calculated as well. The
        rationale behind this approach is that the other fitness criteria available,
        e.g. the null space, do not make more sense if the source estimate is worse.
        In \citet{Liesenborgs5}, an extra mass sheet basis function was introduced to 
        allow for a mass density offset in the strong lensing region. Contrary to the 
        Plummer weights, the weight 
        of this basis function is used directly and does not take part in the rescaling
        procedure just described. The Plummer weights so describe the shape of the
        mass density, on top of an offset described directly by the mass sheet weight.

        The procedure of different refinements of the grid will produce one mass map that
        is considered the best one for this run. Because much randomness is involved
        in the GA itself and random offsets are introduced in the grid
        placements, performing this procedure again will produce a somewhat different
        mass distribution. Therefore, typically several tens of such individual inversion runs 
        are performed, where the average of these solutions will highlight the common features
        while suppressing random fluctuations. The variation between the results of each
        run can provide some insight into the degree to which the mass density in different
        regions is constrained. 
        
        The average as well as the individual models
        are built from Plummer basis functions and are therefore always smooth
        and continuous. Therefore no post-processing needs to be done to visualize
        the resulting mass distribution: the one that is shown corresponds to
        the lens effect that is visible.

    \begin{figure*}
        \centering
        \includegraphics[width=\WF\textwidth]{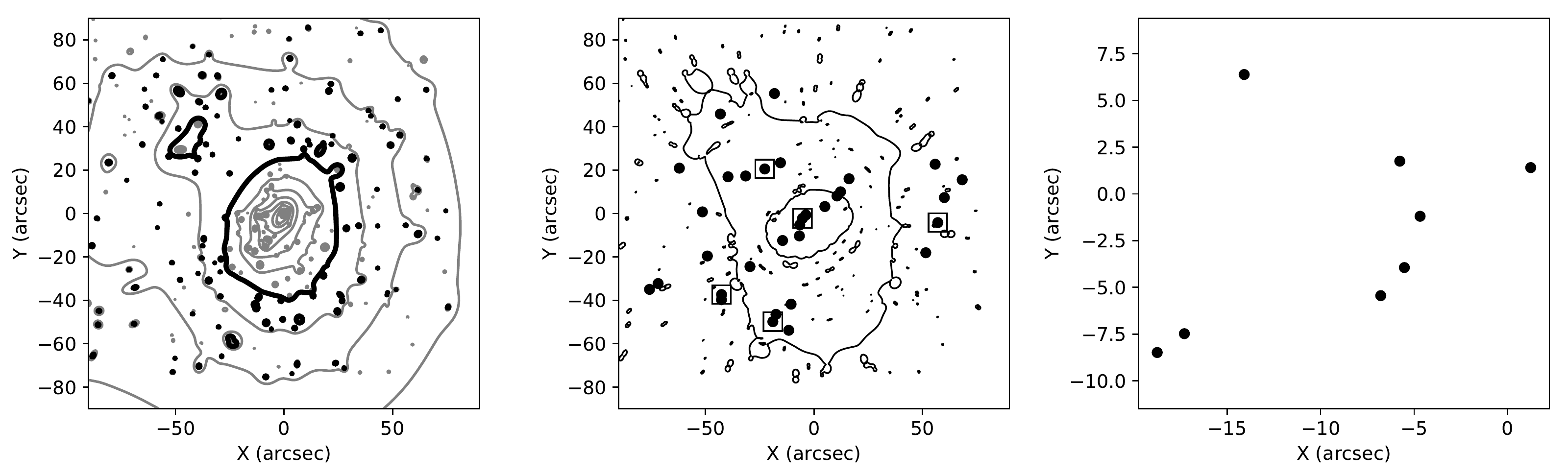}
        \caption{For the time delay test an existing model for
        Abell 1689 was used \citep{2007ApJ...668..643L}, the mass distribution of
        which is depicted by contours in the left panel. There, the solid lines
        indicate $\kappa=1$ (for a source redshift of $z=3$); the spacing between 
        contours is $\Delta\kappa=0.2$. The filled circles in the centre panel
        show the 32 image positions generated by the eight point sources in the
        right panel, the squares indicate the multiple image system
        that is used in the left panel of Fig~\ref{fig:tdrelativetimes}.
        The critical lines shown in the centre panel also refer to
        a redshift of $z=3$.}
        \label{fig:tdreallenspt}
    \end{figure*}

    \subsection{Generalizations}

        The dynamic grid that is used in the inversion strategy serves to fix the
        positions and sizes of the Plummer basis functions, for which the GA will
        subsequently determine the weights. For the GA itself, it is however
        irrelevant that the positions and sizes originate from a grid layout,
        and this has been made explicit: while the user of the inversion code can
        still work with the existing and tried subdivision procedure, there is now
        the option of using a different way to determine this layout of basis functions.
        In sections \ref{sec:substruct} and \ref{sec:weak} this is used to
        facilitate handling both small scale substructure and large scale weak lensing
        measurements.

        In a similar way, the choice of basis function is not relevant to the GA,
        in the sense that once the necessary deflection angles have been calculated,
        it does not matter from which type of basis function they originate. The
        specific choice will certainly have an effect on how well the GA will
        converge to a solution and how well this solution will perform, but it
        does not change the inner workings of the GA. Therefore, the inversion code
        can not only be instructed where to place the basis functions in a more
        flexible way, but the type of basis function can be specified as well. Any
        type that is supported by the \grale{} simulation code can be used here, ranging from
        simple models like a projected Plummer sphere, a square pixel \citep{1998MNRAS.294..734A}
        or a singular isothermal sphere (SIS) over rotated elliptical models to
        even complex composite models. Moreover, different basis functions need
        not originate from the same underlying models, different ones can be used, e.g.
        Plummer basis functions augmented by a few SIS models.
        The use of the mass sheet basis function can be generalized as well: if
        desired, any other supported model can be used instead. While this
        can be a model with a similar effect, e.g. a mass disc instead of a mass
        sheet, this does not need to be the case. Note that the GA still treats this
        type of basis function slightly differently, as its weight is not included in the
        rescaling step mentioned earlier.

    \subsection{Fitness criteria} 

    \begin{figure*}
        \centering
        \subfigure{\includegraphics[width=\WF\textwidth]{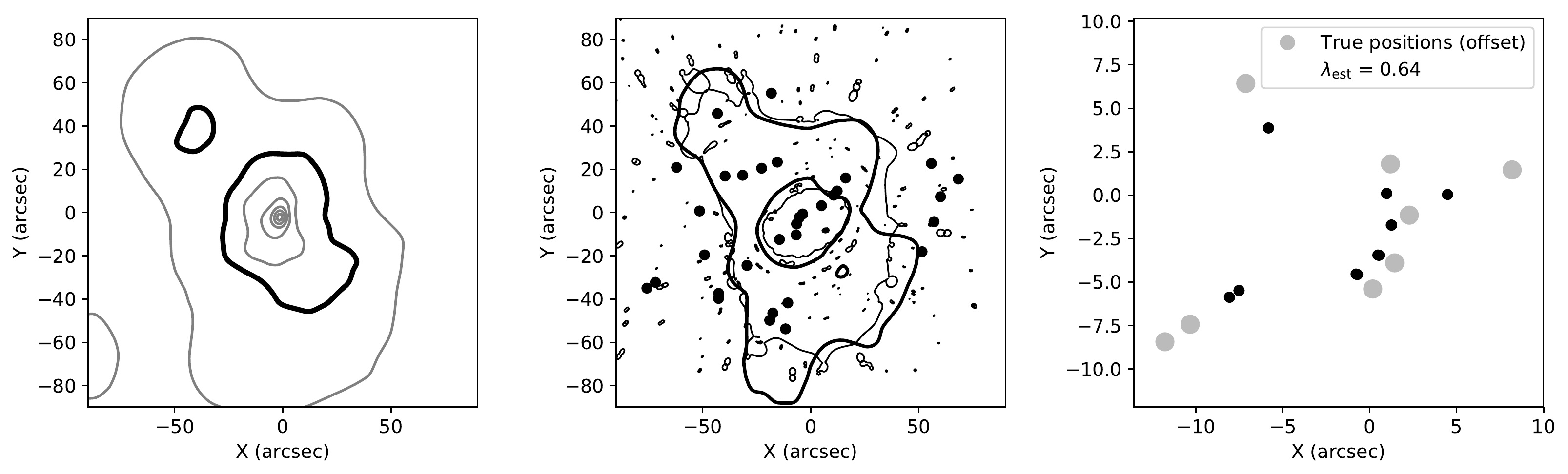}}
        \subfigure{\includegraphics[width=\WF\textwidth]{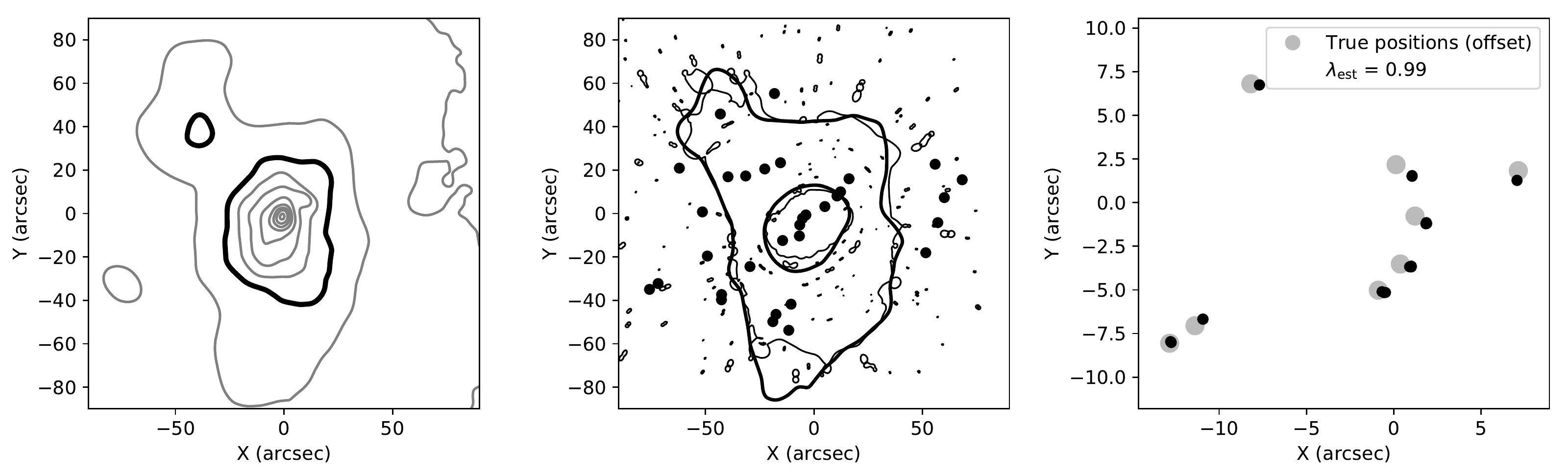}}
        \caption{Results for the time delay test, where the multiple images from
        Fig.~\ref{fig:tdreallenspt} are used as constraints, augmented by time delay
        information for all image locations, as well as null space information. 
        The top half corresponds to the use of $\textrm{fitness}_{\rm TD,2009}$, the
        bottom half to $\textrm{fitness}_{\rm TD,NoSrc}$. The recovered critical
        lines are shown as thick solid lines, the ones of the input lens as thin
        lines for comparison. In the right panels, which show the back-projected images
        (i.e. the estimate of the sources), the true source positions are indicated
        as well, as filled gray circles with a diameter of 1", but offset to have 
        the same centre as the recovered positions. This
        allows one to compare the relative sizes of input and reconstructed source
        planes, thereby visualizing a rescaling due to the MSD. 
        The estimated scale factor $\lambda_{\rm est}$ corresponds to this
        fraction of recovered to true source plane areas, and would quantify the
        relation between the reconstructed and real mass densities according to
        equation (\ref{eq:msd}) in case the exact MSD would
        apply.
        The older time
        delay fitness measure causes over-focusing of the images, while the
        new fitness criterion allows the MSD scale factor be recovered accurately. This MSD 
        difference is also hinted at by the difference in number of density contours,
        indicating that one solution is steeper than the other.
        }
        \label{fig:tdreconstrallpt}
    \end{figure*}

        Central to the optimization procedure are the fitness criteria: each one provides
        a measure of how well a trial solution performs for some specific aspect.
        In the GA that is used in \grale, the precise value of these fitness
        measures does not matter, they are only used when comparing two genomes to
        determine which one is better.  If more than one fitness measure is used,
        roughly speaking, a genome is said to dominate another one if it is better
        with respect to all these fitness values. In a set of genomes one can then identify
        the subset that is not dominated by any other genome: the non-dominated set.
        After excluding this subset, a new non-dominated set can be determined and so on.
        If only a single fitness measure is specified, the trial solutions can be 
        ranked accordingly; if more than one fitness type is used, the population 
        of genomes is subdivided into these non-dominated sets.
        At the core however, one only needs to be able to tell
        which solution is better regarding a single fitness criterion, there is no
        need for differentiability or even continuity of these fitness measures.
        
        In a strong lensing scenario, the central requirement is that different images
        originate from the same source. If extended images are used as input to the inversion
        routine, this requirement is translated into a fitness measure that projects
        the images onto their source planes, and for these back-projected images measures
        how well they overlap. This is done by measuring the distances between the corners
        of rectangles surrounding the back-projected images, and the distances 
        between corresponding image points when available. Such distances are not measured on an
        absolute scale however, instead, the average size of the back-projected images is
        used. As described in \cite{Liesenborgs}, this helps guard against solutions
        that over-focus the images. Note that in this approach differently sized images
        are matched to a single source size, thereby incorporating information about the
        magnification of the images as a whole as well (the magnification of unresolved
        points is not used). For the more complex situation of merging images
        on a critical line, some care should be taken so that these partial source shapes are
        not compared directly to the full source shape from another, complete image. One
        could either use only part of the full image, the part that is visible in the
        merging ones, or, if the images are particularly extended and corresponding points
        can be identified in all images, only use these to determine the overlap and
        not the rectangles.\footnote{Not all corresponding points need to be identified
        in each image.}
        
        The identification of extended images is not always straightforward however,
        and it is actually more common to have point image information
        instead. If this type of input is used, the back-projected image sizes are no longer
        available to base a distance scale on. Instead, as described in \citet{2010MNRAS.408.1916Z},
        the size of the area of all back-projected image points is used. This places some
        control over the GA in the hands of the outermost sources, as they determine this
        scale, which in turn affects the calculated fitness value. If one wants to guard
        against this, it is possible to base this scale on the median absolute deviation (MAD) 
        of the back-projected positions, at the expense of some additional computations.
        In this point image approach, using magnification information of the point 
        images is not available as a constraint in the optimization procedure.
        
        These `overlap' fitness measures work in the source plane, and while the choice of the
        distance scale used for measuring overlap avoids preferring trial solutions that merely
        over-focus the images, in essence it remains a source plane optimization. The 
        back-projected images will not coincide perfectly, and because only a source plane
        criterion is used, there is no guarantee that these differences will stay small in the
        image plane (see also \citet{Kochanek2006}). 
        The root-mean-square (RMS) value of the predicted vs provided image
        positions can be used to measure this, and in practice this has turned out to be
        very acceptable using the fitness measure above. 
        In case a relatively bad RMS is obtained, or if one
        would like to improve it even more, a fitness measure based on
        the differences in the image plane instead of the source plane can be
        used. To do so, these differences in the image plane $\Delta \Vec{\theta}$ are approximated
        by multiplying the corresponding difference in the source plane $\Delta \Vec{\beta}$
        with the inverse of the magnification matrix $\mathcal{A}$ (see equation (\ref{eq:firstorder})),
        similar to the approach in \cite{2010PASJ...62.1017O}, appendix 2.
        Specifically, for a set of corresponding image points, each of the 
        back-projected points in turn is used as a possible source position,
        \footnote{This is the default behaviour, alternatively the average
        of the back-projected image points can be used as well.}
        and differences with other back-projected positions are translated into 
        image plane differences using the magnification matrix. The sum of these squared
        differences is the basis for the fitness measure. To avoid sources with more images
        having a disproportionally large influence during the optimization, per source
        the average of this sum is used. The complete fitness value, for all
        sources, then consists of the sum of these single source contributions.
        By itself this fitness criterion does not seem to yield source plane points that
        overlap as well as the other overlap fitness measures, but using both fitness measures
        together in the multi-objective GA usually provides solutions which perform well on both
        accounts -- at the expense of an extra fitness criterion however.

        Depending on the number of observed images that can be used for the lens inversion,
        it may be possible for the GA to evolve towards a solution that, while correctly
        predicting the input images, also predicts unobserved extra images. To help the
        GA steer clear of such sub-optimal solutions, an additional null space fitness
        measure can be specified. For a source with extended images, one typically creates 
        a grid of triangles encompassing a part of the image plane, where not only regions 
        containing the observed images are cut out, but also regions where unobserved images
        are possible, e.g. behind bright cluster galaxies. By projecting these triangles onto
        the source plane and determining the amount of overlap with the estimated source,
        a value is obtained that expresses if there are extra images and how prominent they
        are \citep{Liesenborgs2}. For point images, a more straightforward approach is used:
        there, a simple uniform grid of triangles is used, i.e. no regions are cut out, and
        the number of back-projected triangles that overlap with the estimated source are
        counted, thereby providing a rough estimate of the number of images of the source
        \citep{2010MNRAS.408.1916Z}. For both approaches, the grids have typically been
        based on a uniform subdivision of an image plane area ranging from $48\times 48$ 
        to $64\times 64$ square grid cells, each cell consisting of two triangles.
        Specific regions, e.g. the observed images, can be removed from this
        uniform grid automatically.
        The grid is taken to be larger than the area of the images themselves to avoid failing
        to detect extra images that lie farther away from the central region -- which would not
        overlap with any of the triangles.

        In \citet{Liesenborgs5}, a fitness measure was described in case time delay
        information is available, to study the constraints provided in the
        SDSS~J1004+4112 lensing system. In section \ref{sec:timedelays} we shall introduce an
        alternative formulation and investigate the performance of the existing and new
        fitness measures. 

        \begin{figure*}
            \centering
            \includegraphics[width=\WF\textwidth]{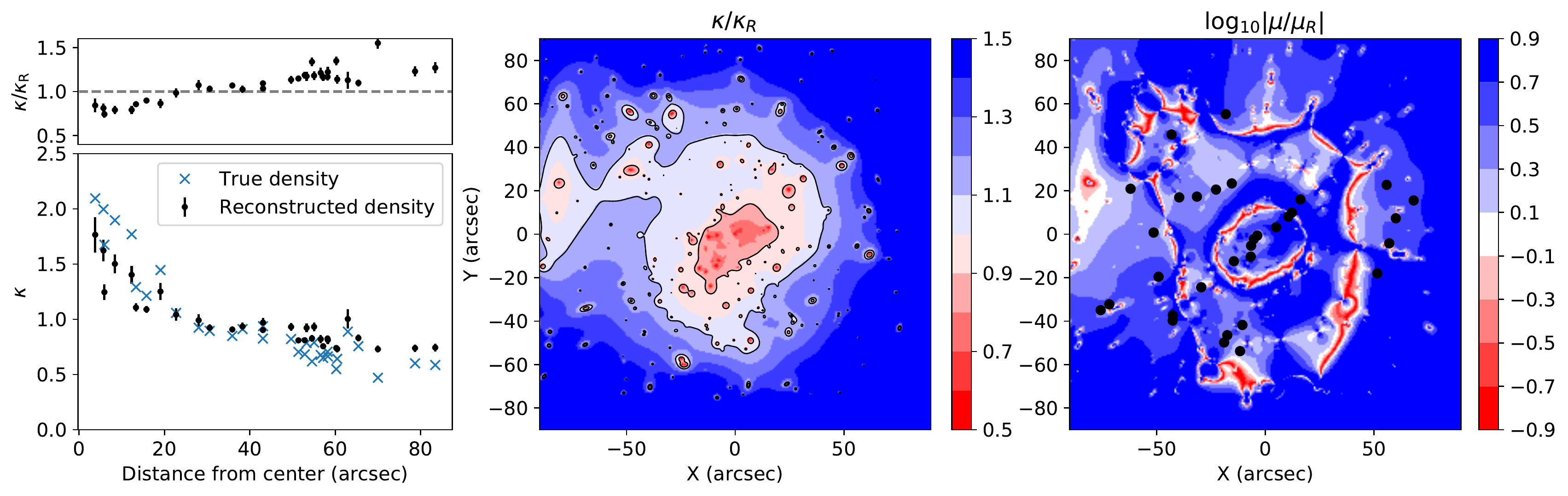}
            \includegraphics[width=\WF\textwidth]{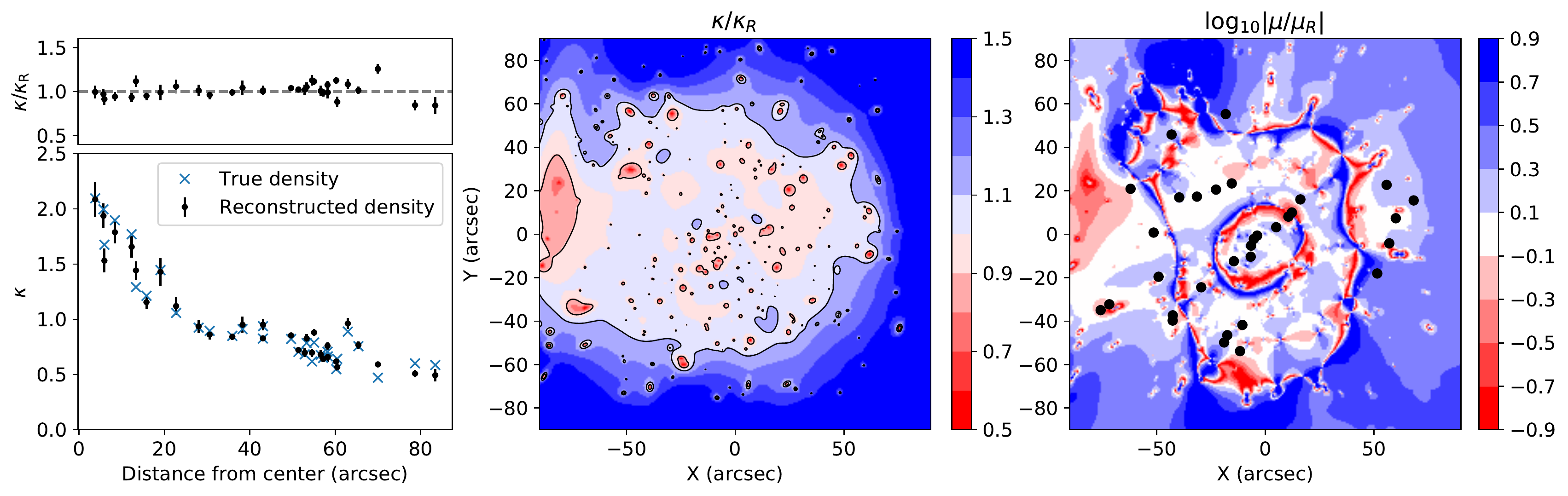}
            \caption{
            In each of the three panels, the top figure corresponds to the solution
            obtained with $\textrm{fitness}_{\rm TD,2009}$ and the bottom figure to
            the one obtained with $\textrm{fitness}_{\rm TD,NoSrc}$.
            Left panel:~the larger plots show both input and recovered densities
            measured at the positions of the images, while the smaller ones show
            the fraction of the recovered density $\kappa$ to the real density
            $\kappa_R$. For visualisation purposes, only the
            distance of a point to the coordinate centre is shown on the horizontal axis.
            Whereas the newly proposed fitness measure yields densities that match the
            true ones well, the other one produces solutions which are
            less steep, and which have a more prominent density offset, i.e. that differ
            by the MSD.
            Centre panel:~these maps show the $\kappa/\kappa_R$ fraction in the
            region under consideration. The solid black line marks the 10\%
            boundaries. While the $\textrm{fitness}_{\rm TD,2009}$
            solution has clear difference in both the central and outer regions,
            similar as what can be seen in the left panel, the solution obtained
            with the new fitness measure is well constrained over
            a large area.
            Right panel:~these plots compare the magnification of the recovered
            lens model $\mu$ to the true magnification $\mu_R$, both for a redshift of $z=9$.
            The consistently larger magnification for the older fitness measure
            is again a sign of the MSD. The new $\textrm{fitness}_{\rm TD,NoSrc}$
            on the other hand produces a much more consistent map.
            }
            \label{fig:tdrelativequant}
        \end{figure*}

        While \grale{} is designed for the inversion of strong lensing systems, having
        information about the larger, weak lensing region is becoming increasingly common.
        It would therefore be desirable to be able to integrate the available weak lensing
        measurements into the inversion procedure. Section \ref{sec:weak} formulates a
        fitness measure that can be added to the multi-objective GA, and investigates
        the information that can be retrieved in this way, for various degrees of correctness
        of the weak lensing shear estimates.

\section{Time delays}\label{sec:timedelays}
    
    \begin{figure*}
        \centering
        \includegraphics[height=0.4\textheight]{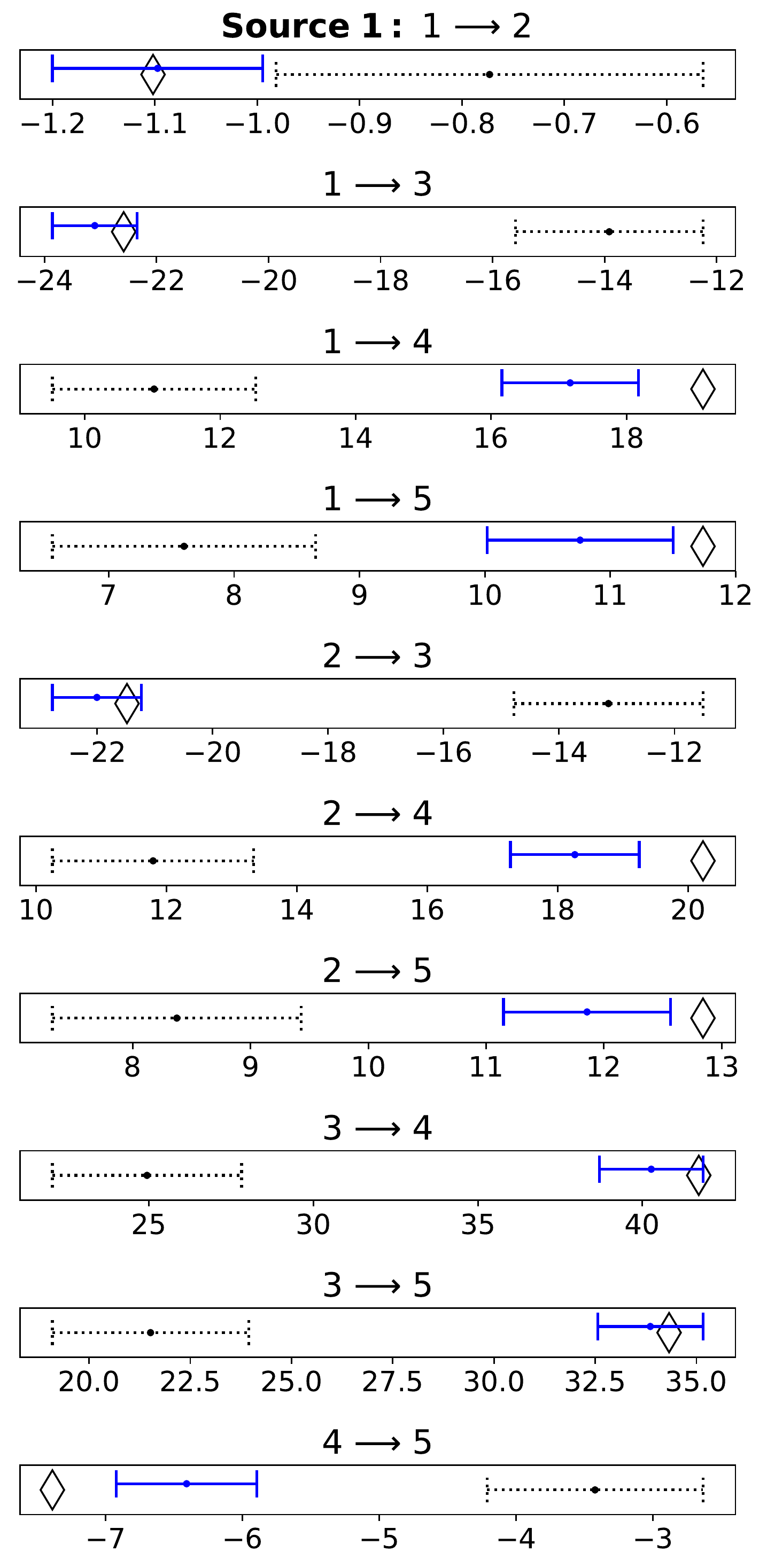}
        \includegraphics[height=0.4\textheight]{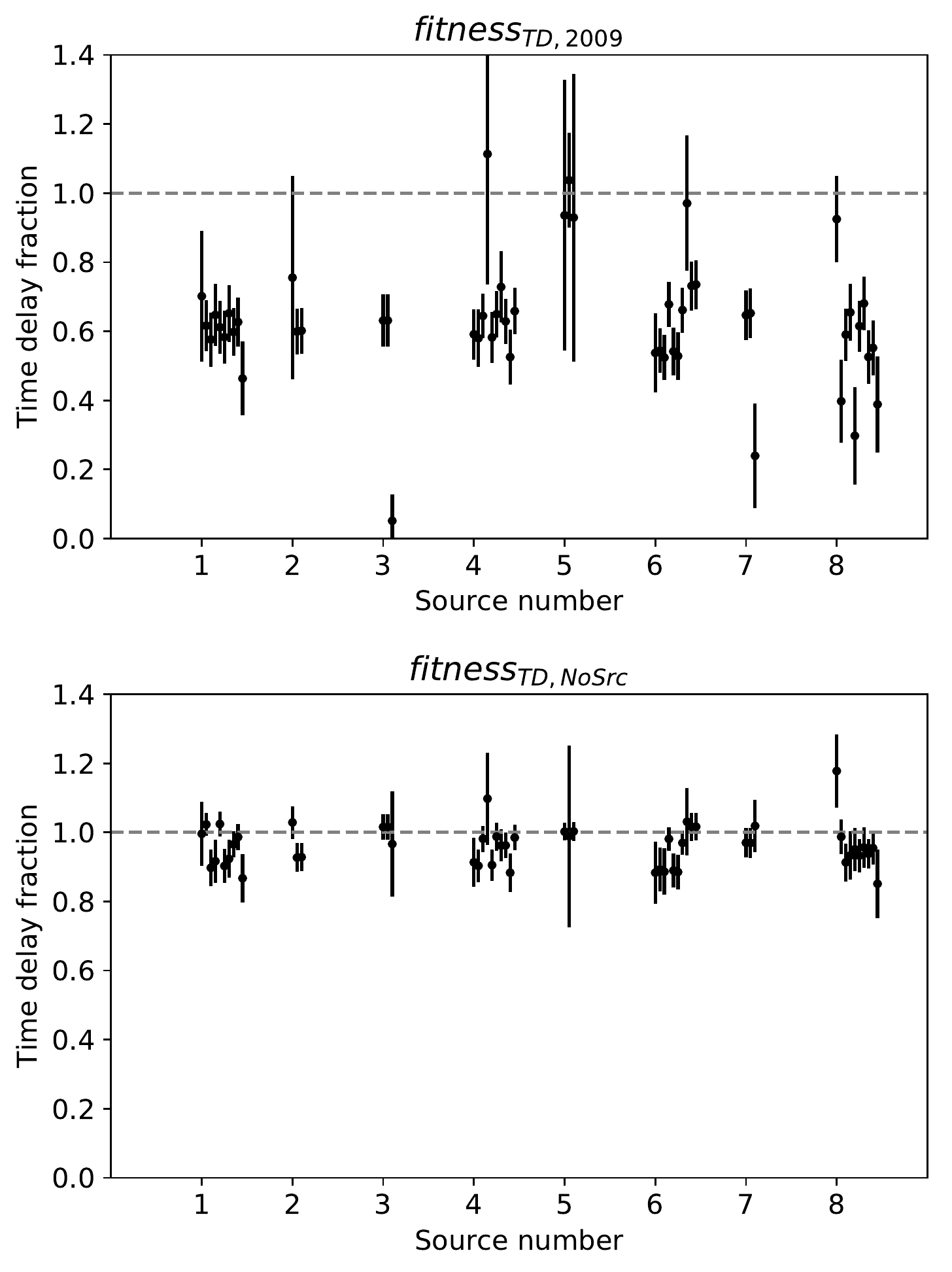}
        \caption{
        Left panel:~this figure compares the predicted time delays
        to the input ones (diamond symbol) for all image pairs of one
        particular source, indicated by squares in the centre panel
        of Fig.~\ref{fig:tdreallenspt}; the time delays are shown as a number of days.
        The dotted black line shows the average and standard deviation
        when the existing $\textrm{fitness}_{\rm TD,2009}$ is used, whereas the solid
        blue lines correspond to the new $\textrm{fitness}_{\rm TD,NoSrc}$.
        For this system, as well as the ones that are not shown in the
        figure, the latter fitness measure produces time delay predictions that 
        correspond better to the input time delays.
        Right panels:~instead of showing the time delays themselves for all
        image pairs, these figures show the fraction of predicted time delay
        to true time delay for all sources, and for all image pairs per
        source. The error bars again indicate averages and standard deviations.
        Clearly, the new fitness measure provides much better correspondence
        with the true time delays. The fact that $\textrm{fitness}_{\rm TD,2009}$
        produces time delays that are consistently shorter can be largely
        attributed to the MSD, which would scale time delays by a certain
        factor.
        }
        \label{fig:tdrelativetimes}
    \end{figure*}

    When available, time delay measurements between images of the same source
    provide especially valuable information. As shown in equation (\ref{eq:timedelay}),
    they directly probe (non-local) differences of the projected potential.
    The image positions themselves only provide information about its 
    gradient (equation (\ref{eq:gradpsi})), and local image deformations 
    even only sample the
    curvature of the projected potential. As explicitly demonstrated in
	\citet{2012MNRAS.425.1772L}, time delay information is therefore very
    useful in breaking the MSD.

    The original fitness measure for including time delay information is based 
    directly on (\ref{eq:timedelay}).
    That equation mentions a single source position however, which is not
    known at optimization time, and for this reason each of the back-projected
    image points $\Vec{\beta}_k = \Vec{\beta}(\Vec{\theta}_k)$ is used
    as a possible source position. Calling
    \begin{equation}
        \Delta t_{ij,kl} \equiv t(\Vec{\theta}_i, \Vec{\beta}_k) - t(\Vec{\theta}_j, \Vec{\beta}_l) \mcm
    \end{equation}
    the time delay fitness contribution for a single source was given by
    \begin{equation}
        \textrm{fitness}_{\rm TD,2009} = \sum_{i \in T} \sum_{\substack{j \in T \\ j \ne i }} \sum_{k=1}^N \sum_{l=1}^N
             \left(\frac{\Delta t_{ij,kl} - \Delta t_{{\rm obs}, ij}}{S_{ij}}\right)^2 \mcm
    \end{equation}
    where the subscript 2009 refers to \citet{Liesenborgs5},
    where this equation was introduced.
    The set $T$ is the subset of image positions for which a time delay
    measurement exists, $N$ is the number of images of the source under
    consideration, and in case multiple sources with time delays are available
    these contributions are simply added.
    The comparison scale $S_{ij}$ is by default set to $\Delta t_{{\rm obs},ij}$,
    administering all time delays the same relative importance. As this may
    be impractical in case a time delay is close to zero for example,
    a different scale value may be set instead\footnote{Currently only a single
    value for all time delays for a source can be set.}.

    Remembering that this fitness measure will be used in conjunction with
    the default positional fitness, the effort to compensate for the unknown
    source position in the expression above actually appears to incorporate
    this overlap requirement as well. Interestingly it has been shown in e.g. 
    \citet{1984A&A...141..318B} and \citet{1988ApJ...327..693G} that the source
    position can be eliminated from the time delay expression: expanding the expression
    for $\Delta t_{ij}$ and noting that 
    $\Vec{\beta} = \Vec{\theta}_i - \Vec{\alpha}_i = \Vec{\theta}_j - \Vec{\alpha}_j$,
    one finds
    \begin{equation}
        \Delta t_{ij} \propto 
            \frac{1}{2}(\Vec{\theta}_i - \Vec{\theta}_j)\cdot(\Vec{\alpha}_i + \Vec{\alpha}_j)
                  - \psi(\Vec{\theta}_i) + \psi(\Vec{\theta}_j) \mcm
    \end{equation}
    where $\Vec{\alpha}_i \equiv \Vec{\alpha}(\Vec{\theta}_i)$ and the proportionality
    factor is the same as in (\ref{eq:timedelay}). Comparable to the fitness
    measure above, we can now base the time delay fitness on
    \begin{equation}
        \textrm{fitness}_{\rm TD,NoSrc} = \sum_{i \in T} \sum_{\substack{j \in T \\ j \ne i }}
             \left(\frac{\Delta t_{ij} - \Delta t_{{\rm obs}, ij}}{S_{ij}}\right)^2 \mpt
    \end{equation}

    \begin{figure*}
        \centering
        \includegraphics[width=\WF\textwidth]{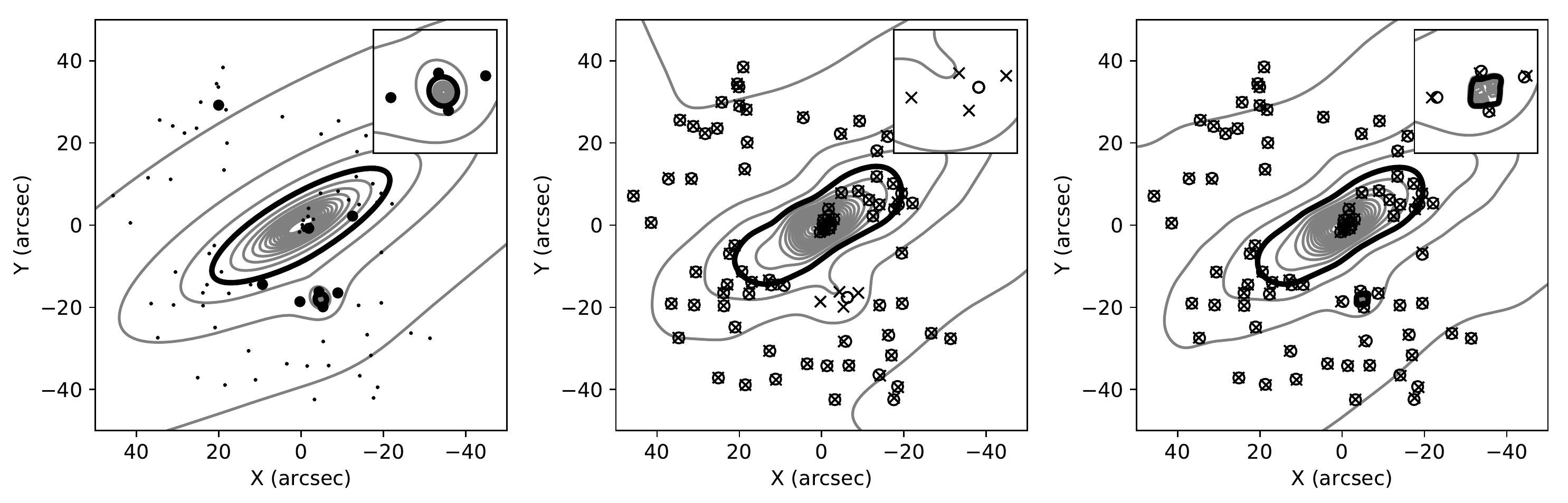}
        \caption{Left panel:~the mass distribution used to study the situation 
        where small scale substructure causes an extra quad to appear. Both
        large scale distribution and small perturbation are modelled using
        NSIE lenses, placed at $z=0.4$. The contours indicate lines of equal 
        convergence $\kappa$, calculated for a redshift of $z=1.5$, which is the redshift
        of the multiple image system containing the quad (filled circles); 
        the inset shows this region in more detail. The thick solid line indicates
        $\kappa=1$, the spacing between contours corresponds to $\Delta\kappa = 0.25$.
        Small points indicate the other point images.
        Centre panel:~the default subdivision procedure leads to this reconstruction,
        an average of 20 individual reconstructions. The input point images are
        shown as crosses, the corresponding images predicted by the model as open
        circles. As shown in the inset, the quad images cannot be reproduced, as
        the required density perturbation is not recovered.
        Right panel:~similar to the centre panel, but this time extra basis
        functions were used to allow for a smaller resolution in between the
        quad images (see text). This time, the quad can indeed be reproduced.
        }
        \label{fig:smallexpmasses}
    \end{figure*}

    Below we shall explore the relative performance of these two fitness measures, where
    the inversions were performed using the default grid-based approach
    to lay out the Plummer basis functions, and with a mass sheet basis function enabled.
    Apart from the positional and time delay fitness measures, the null space criterion
    was used as well. The lens model used is the \lenstool{} one
    for Abell 1689\footnote{Available from the \lenstool{} web site: \url{https://projets.lam.fr/projects/lenstool/wiki}} 
    \citep{2007ApJ...668..643L}, which is merely used as a
    an example of an underlying true mass density, i.e. the simulated images have no relation with the
    images observed in that cluster. Instead, 
    eight source positions were chosen, generating 32 images.
    To assess the effect of adding the time delay fitness measures,
    in these tests all of the multiple image systems were equipped with
    time delay information. 

    Neither in this test, nor in the ones in the next sections,
    the true lens model originates from an N-body simulation. The non-parametric
    inversion method uses a multitude of basis functions to be able to model
    mass distributions with rather arbitrary shapes, implying that the origin
    of the observed images should not matter. The images can be based on simulations
    using an analytical model, as in this example, using more complex N-body
    simulations, e.g. as in \citet{2017MNRAS.472.3177M}, or, 
    of course, real world observations. To assess properties of the reconstruction
    procedure in general, one would not want to be restricted by the kinds of models
    that N-body simulations produce, but allow more flexible mass distributions,
    e.g. differing by an MSD-like scale factor, as well.

    The true lens as well as the generated images and corresponding sources 
    can be seen in Fig.~\ref{fig:tdreallenspt}. The reconstructions for this 
    system, for both fitness measures, are depicted in Fig.~\ref{fig:tdreconstrallpt}.
    In the top half, where $\textrm{fitness}_{\rm TD,2009}$ was used, the contours 
    of the recovered density indicate a distribution that is less steep, and has a
    larger density offset than in the bottom half, for $\textrm{fitness}_{\rm TD,NoSrc}$.
    A similar effect can be seen when comparing both reconstructions in
    the left panel of Fig.~\ref{fig:tdrelativequant},
    where the densities at the locations of the images are shown, and is
    highly suggestive of the presence of the MSD. The centre panel of the
    same figure shows the fraction of the recovered density $\kappa$ to the
    real density $\kappa_R$, in the part of the lens plane under consideration.
    Whereas the recovered density using $\textrm{fitness}_{\rm TD,2009}$ only
    lies within 10\% of $\kappa_R$ in a relatively small region, the correspondence
    for the newly proposed fitness measure clearly covers a much larger region,
    with main differences where the mass peaks in the true model are located.

    This MSD effect can also be seen in the right panel, where the magnification
    is shown for a source at a redshift of $z=9$, similar to figs 21 and 22
    of \citet{2017MNRAS.472.3177M}. Note that this redshift
    dependence causes the actual magnification of the
    images, which correspond to other redshifts, to differ. As the magnification
    can change over orders of magnitude, we have chosen to plot the logarithm.
    The consistently different magnification for the old fitness measure is
    again a manifestation of the MSD, which is far less problematic with
    the new fitness measure. Still, being dependent on second derivatives
    of the projected potential, the magnification can be rather sensitive to
    small differences in models. For the new fitness measure, differences
    can therefore also be seen, although far less consistently, and most
    pronounced where there are no image constraints.

    In the right panels of
    Fig.~\ref{fig:tdreconstrallpt}, the recovered sources are compared to the input
    source positions, where the latter are offset to have the same centre location.
    In a gravitational lensing scenario, the overall offset of the source 
    positions cannot be constrained, as is explained in more detail in
    appendix~\ref{app:srcpos}. By using this offset in
    the plots, the scales of the recovered vs. input source planes can 
    however easily be compared visually.
    The exact MSD from equation (\ref{eq:msd}) as well as the
    generalized versions cause differently scaled source planes to
    correspond to the same observed images, and the estimated $\lambda_{\rm est}$
    from these recovered and real source plane scales can therefore be used
    to indicate the degree to which the correct solution has been retrieved.
    Note that the relation between source sizes and image
    sizes is precisely what the magnification corresponds to, so the 
    consistently larger magnification in the top-right part of
    Fig.~\ref{fig:tdrelativequant} is to be expected based
    on this difference in source plane scales.

    \begin{figure*}
        \centering
        \includegraphics[width=\WF\textwidth]{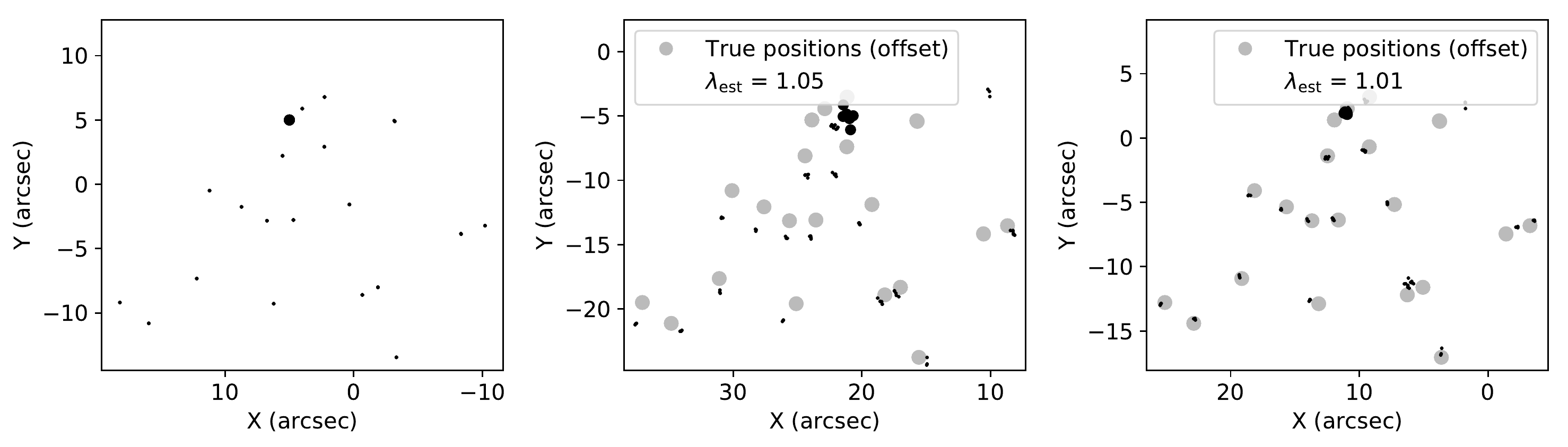}
        \caption{Left panel:~the 20 true source positions, that generate the 
        multiple image systems shown in the left panel of Fig.~\ref{fig:smallexpmasses}.
        Here too, the source responsible for the quad is shown as a filled
        circle.
        Centre panel:~the back-projected images using the reconstruction in
        the centre panel of the same figure, compared to the true source positions;
        the diameter of the gray circles is 1". 
        Similar to a previous example, the true source positions are shown
        with an offset, to be able to compare true and recovered source plane
        scales. While the overall scale is recovered well, as are most sources,
        especially the images of the quad system do not coincide satisfactorily.
        Right panel:~same, but using the reconstruction of the right panel of
        Fig.~\ref{fig:smallexpmasses}. The agreement of the source plane scales
        is even more similar, but especially noticeable is the fact that the
        points of the quad system now overlap to a much better degree.
        }
        \label{fig:smallexpbp}
    \end{figure*}

    Based on these results, using the newly proposed fitness measure appears 
    to be advantageous, as both densities at the images positions and scale 
    of the source planes are recovered more accurately.
    To further assess the performance of the two time delay fitness measures,
    the time delays that were provided as input will be compared to the predicted time
    delays, which are calculated as follows. First, the image points for the same source
    are projected back onto their source plane and 
    their average -- the straightforward average, not using weights based on the
    magnification -- is used as the source position $\Vec{\beta}$.
    The image positions 
    $\Vec{\theta}$ that correspond to this $\Vec{\beta}$ are recalculated, yielding image
    position predictions that differ somewhat from the input positions. This $\Vec{\beta}$ and
    these $\Vec{\theta}$ positions are subsequently used to calculate time delay
    differences using equation ($\ref{eq:timedelay}$), and finally compared to the input
    time delays. In this experiment 40 individual solutions were generated to estimate
    accuracy and precision.

    The left panel of Fig.~\ref{fig:tdrelativetimes} shows 
    the averages and standard deviations, based on the 40 individual solutions,
    of the time delay predictions for this test, for the images of a single source.
    The dotted black line shows these quantities for the existing fitness measure, 
    whereas the solid blue line shows those for the newly proposed one. 
    Plots for the other sources show the same trends, which can in fact
    be more clearly seen in the other parts of the figure, where for each source, and each
    image pair, the time delay fraction $\Delta t_{ij}/\Delta t_{ij,{\rm obs}}$
    is shown. The quite consistently smaller time delay sizes for
    $\textrm{fitness}_{\rm TD,2009}$ can again be seen as an MSD effect,
    rescaling the projected potential as well as the density.
    For this test, the effect is quite clear: the new fitness measure which 
    eliminates the unknown source position produces predictions that match the 
    input time delays to a much better degree.

    \begin{figure*}
        \centering
        \includegraphics[width=\WF\textwidth]{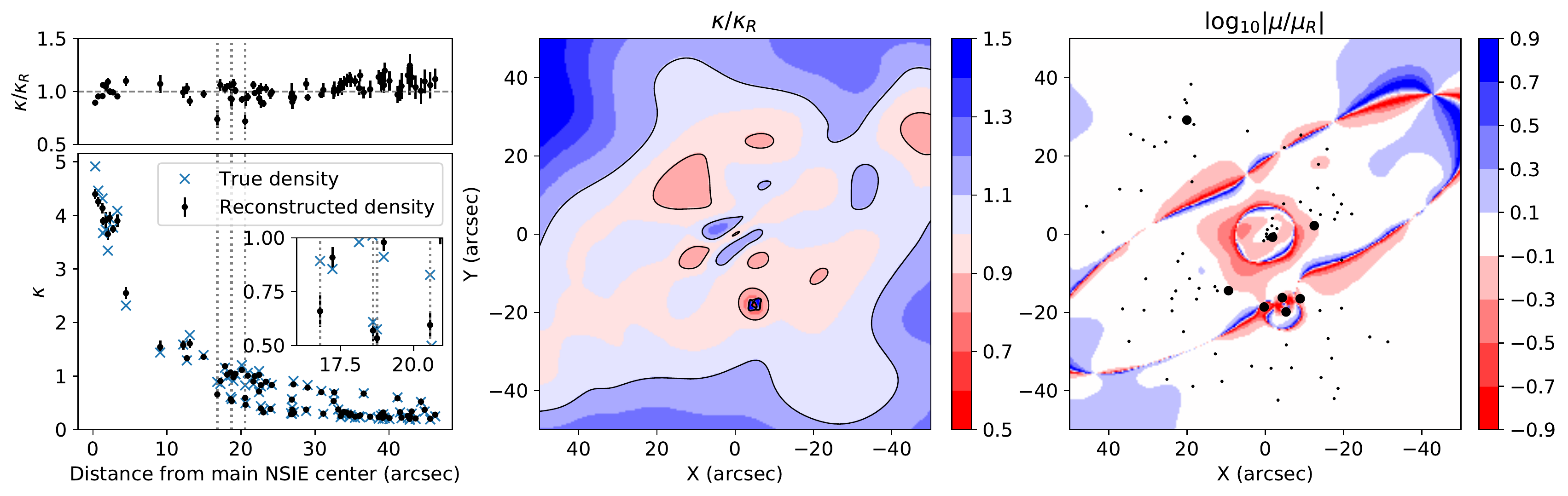}
        \caption{
        Left panel:~using the recovered mass map shown in the right panel 
        of Fig.~\ref{fig:smallexpmasses}, the densities $\kappa$
        at the positions of the images are compared to the densities of the real lens,
        $\kappa_R$; the positions of the quad images are indicated by vertical lines,
        more clearly discernible in the inset. While there are certainly differences, 
        the overall steepness as well as the density offset appear to be recovered 
        correctly.
        Centre panel:~the relative density, $\kappa/\kappa_R$, of recovered and true
        lens models, over the lens plane region; the black line again indicates a
        change of 10\%. This boundary covers a large part of the region, with a
        notable exception at the position of the quad, indicating that further, more
        local degeneracies can play a role.
        Right panel:~comparison of the recovered magnification $\mu$ to the one of
        the real lens model, $\mu_R$. The correspondence shows much less variation
        as compared to the example in Fig.~\ref{fig:tdrelativequant},
        undoubtedly due to a less complex mass model needing to be recovered.
        }
        \label{fig:smallexprelquant}
    \end{figure*}

    \begin{figure*}
        \centering
        \includegraphics[width=\WF\textwidth]{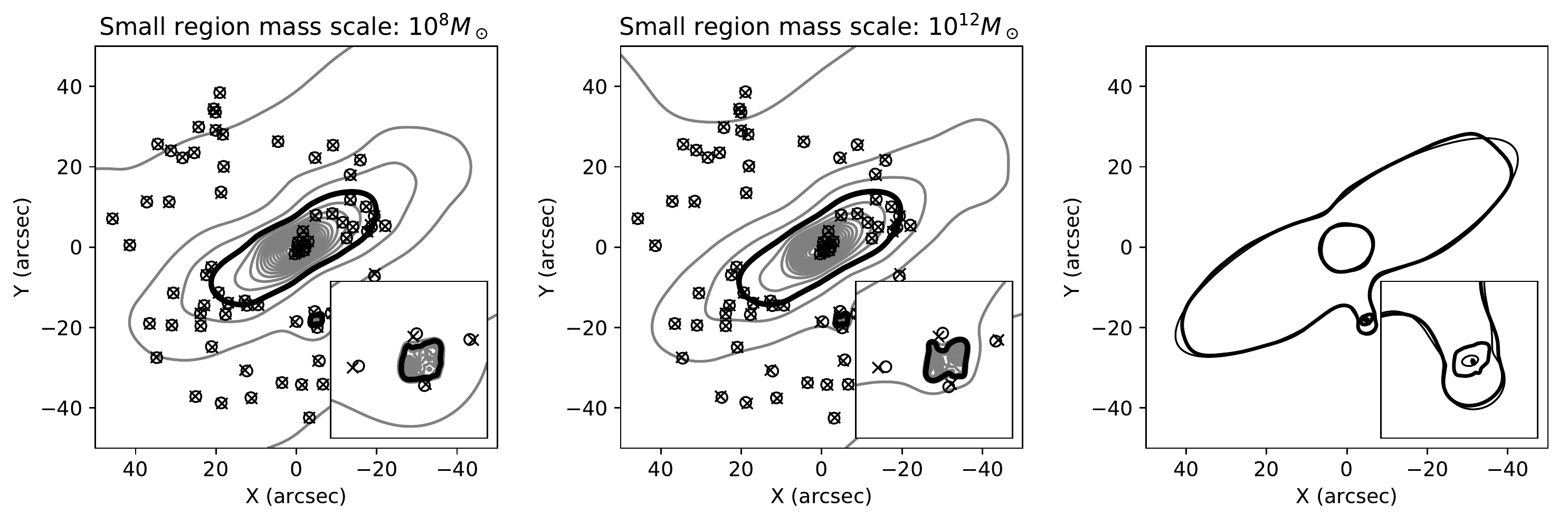}
        \caption{The first two panels show results similar to the right panel
        of Fig.~\ref{fig:smallexpmasses}, but for different settings of the mass
        scale in the region of the small density peak (see text). The right panel
        shows the critical lines at the redshift $z=1.5$ of the source, for both
        the reconstruction from the right panel of Fig.~\ref{fig:smallexpmasses}, 
        and the true lens.
        }
        \label{fig:smallexpmassscales}
    \end{figure*}

    \begin{figure}
        \centering
        \includegraphics[width=\WFTHREE\textwidth]{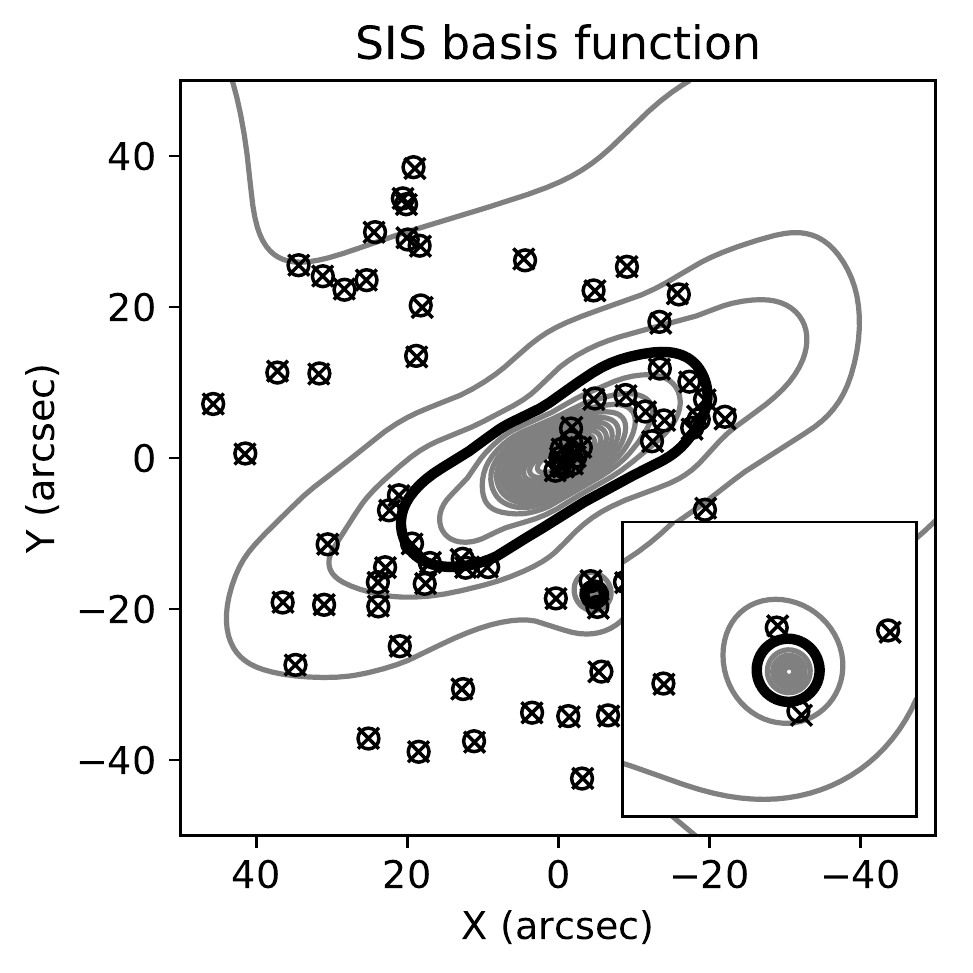}
        \caption{Instead of using a grid of Plummer basis functions to account for the small
        scale substructure, one can imagine that based on the visible light a single
        basis function is used. In this example, a SIS lens was placed at 
        the centre of the mass peak inside the quad, and its weight
        was subsequently optimized during the inversion procedure. As shown, the resulting
        model is also capable of explaining the observed images, including the quad. 
        For this inversion to succeed consistently, the extra RMS fitness measure 
        needed to be enabled.
        }
        \label{fig:smallexpsis}
    \end{figure}

    \begin{table*}
        \centering
        \begin{tabular}{l|cccccccccccccccccccc}
            Source&1&2&3&4&5&6&7&8&9&10 \\
            \hline
            Default RMS &0.17 &0.22 &0.16 &0.25 &0.27 &0.12 &0.11 &0.14 &0.17 &0.067\\
            Substructure RMS &0.24 &0.19 &0.15 &0.26 &0.20 &0.15 &0.092 &0.14 &0.15 &0.078\\
            \ \\
            Source&11&12&13&14&15&16&17&18&19&{\bf 20}\\
            \hline
            Default RMS&0.32&0.41&0.23&0.14&0.19&0.11&0.25&0.10&0.076&{\bf 2.8}\\
            Substructure RMS&0.32&0.42&0.18&0.17&0.14&0.081&0.15&0.17&0.061&{\bf 0.22}\\
        \end{tabular}
        \caption{For each of the sources used in the substructure test,
        the RMS comparing the true image locations to the predicted ones is specified in 
        arc seconds. The first line, ``Default RMS'' shows the RMS that results from the default procedure,
        corresponding to the reconstruction in the centre panel of Fig.~\ref{fig:smallexpmasses}.
        The second line, ``Substructure RMS'', are the values for the model
        in the right panel of the same figure. Only source 20, the one containing the 
        small scale quad, shows a clear difference.}
        \label{tab:rmssubstruct}
    \end{table*}
\section{Substructure}\label{sec:substruct}

    The default procedure which uses the subdivision grid to arrange basis functions
    can have problems recovering small scale substructure: for basis
    functions with the required resolution to be present, one would need the
    mass threshold responsible for splitting a grid cell into four new cells to be relatively
    low, as such small scale substructure would not enclose a particularly
    massive region. This in turn causes other regions to be subdivided quite
    finely as well, leading to a very large number of basis functions. Without
    an adequate number of available constraints, the GA would easily evolve
    to a sub-optimal solution, essentially getting lost in the parameter space.

    As mentioned in \cite{2019MNRAS.482.5666W}, this insufficient resolution was
    the case in the inversion of MACS~J1149.6+2223. In that cluster, a well
    resolved background galaxy can be seen as three separate large images,
    and a supernova, SN Refsdal \citep{2015Sci...347.1123K},
    in one of the spiral arms was actually visible four times in one of 
    these three images.
    This quad, with a relatively small separation, is generated by
    a cluster galaxy that overlaps with one of the larger images.
    The increased flexibility for placing basis functions that was described
    earlier, allows one to handle such cases with
    small scale substructure, where one has a strong indication that extra
    mass needs to be present at a particular location, for example because
    of the presence of a cluster galaxy as in MACS~J1149, combined with a
    lack of accuracy in the initial reconstruction. While it
    has now become easy to add small scale substructure throughout the lensing
    region, in our opinion one should only make use of this when the default
    inversion procedure fails to recover something fundamental, e.g. the
    multiplicity of a lensed source.

    To study a similar situation, the simulated gravitational lens shown
    in the left panel of Fig.~\ref{fig:smallexpmasses} was used: the overall
    elliptical mass distribution at $z=0.4$ causes 81 images of 20 point sources, where four
    of the images are created by the presence of a carefully placed small mass clump at
    $(-4.8", -18.1")$, also shown in the inset. For both the main component and the small
    perturbation a non-singular isothermal ellipse (NSIE) model was used.
    The true source positions can be seen in the left panel of Fig.~\ref{fig:smallexpbp}.

    The centre panel of Fig.~\ref{fig:smallexpmasses} shows the recovered mass distribution,
    an average of 20 individual solutions, when the default procedure using
    the dynamic subdivision grid is used. While the solution does hint at the
    presence of extra mass in between the quad images, a comparison of the input image
    positions (crosses) to the predicted image positions (circles) shows that
    the reconstruction does not have the required resolution to predict all images
    of the quad -- in fact, only a single image position is recovered. The
    back-projected image positions, i.e. the estimated source positions, are
    shown in the centre panel of Fig.~\ref{fig:smallexpbp}. While the overall
    scale of the situation is very similar to the one for the input sources,
    as can be seen from the $\lambda_{\rm est}$ value which expresses
    the fraction of these scales, and which indicates that the MSD scale factor and
    offset are recovered well, the back-projected images of the 
    system containing the quad clearly overlap less well than the other ones.

    To allow this source and the quad images to be recovered more accurately, in a small region
    in between these images other basis functions were added. Overall, the same subdivision
    steps were used as in the default inversions, but at each subdivision step
    basis functions based on a small, uniform $15\times15$ grid were added
    as well. Given the paucity of constraints for the quad region, these extra
    basis functions should provide more than adequate flexibility. The result of this
    slightly modified procedure can be seen in the right panel of Fig.~\ref{fig:smallexpmasses}.
    Thanks to the extra basis functions available in between the quad images,
    mass could be placed there allowing the reconstructed lens to
    predict the existence of these images as well. 
    
    In the right panel of Fig.~\ref{fig:smallexpbp}
    the back-projected images are shown for this reconstruction, showing that
    the images of the system with the quad now overlap to a much better degree.
    The overall scale of the recovered sources still corresponds very well to
    that of the true sources, indicating that the MSD scale factor is found accurately. This is
    further supported by the left panel of Fig.~\ref{fig:smallexprelquant} that compares the mass
    densities of the input lens to this reconstruction. While there certainly are
    differences, mostly near the main mass peak, the overall steepness appears
    very similar, as does the density offset. In fact, the similarity
    is within 10\% of the true mass density for a considerable part of the
    lens plane area, as can be seen in the centre panel of the figure. The
    right panel compares the magnifications of the recovered and true densities,
    indicating a correspondence that could be expected from the matching
    source plane scales. 

    When comparing all back-projected images to the true sources, it does appear
    that the reconstruction in the right panel of Fig.~\ref{fig:smallexpbp} performs
    in general better than the one without the small scale basis functions in 
    the centre panel. The true source positions are never an observable however,
    and as described in appendix~\ref{app:srcpos} cannot be fixed. The only
    true criteria one can use to ascertain how well a reconstructed model
    fits the observations are how well the back-projected images correspond
    to a single source, and more importantly how well the re-traced images
    from the estimated source positions correspond to the observed images. For
    all multiple image systems except for the quad, both solutions, with and
    without the small scale substructure, perform similarly. This
    can be seen explicitly in Table~\ref{tab:rmssubstruct},
    where the RMS for each system has been calculated in the following way:
    the average of the back-projected images is used as the source position,
    and starting from the observed image positions, these are modified to
    correspond to that source position. The table shows a clear difference for
    source 20, the one containing the quad, while the images of the other
    sources all have a very similar RMS in the two reconstructions.

    For each set of basis functions, the GA will look for appropriate weights.
    The initial weights, e.g. the initial masses of the Plummer basis functions,
    as well as their relative contributions will of course affect this search.
    For the overall mass distribution, the total mass required is automatically estimated
    from the separations in the multiple image systems, and the initial weights
    are chosen to correspond to this total mass. For the extra basis functions,
    which are needed to reproduce the quad images, no such automated procedure
    is provided however, and an extra mass scale for this region needs to be
    provided manually. In the reconstruction shown in the left panel of Fig.~\ref{fig:smallexpmasses},
    this mass scale was set to $10^{10}M_\odot$.

    Fortunately, the procedure does not appear to be very sensitive to the
    choice of this initial mass scale, as can be seen in Fig.~\ref{fig:smallexpmassscales}.
    There, similar reconstructions are shown, only differing in this mass
    scale for the small region, changing over four orders of magnitude, from
    $10^8 M_\odot$ to $10^{12} M_\odot$. The details are different, as there
    are only very few constraints for the small region, but in each case a
    reconstruction could be found. The addition of a small null space constraint,
    only for the region around the quad, was helpful in preventing the 
    optimization from placing too much mass there, mainly for the larger mass
    scales. The right-most panel 
    compares the recovered critical lines to the ones of the true
    lens. While there are some differences, which can be expected due to
    the lack of constraints in the region of the small mass density peak,
    the overall correspondence is good.

    Instead of using a grid of Plummer basis functions to be able to account for
    a wide variety of mass distributions in between the quad images, one could imagine 
    using a single, more simple density profile. To illustrate
    that the extensions to \grale{} now make it possible to combine the
    default, Plummer basis functions with a different one, here we
    use a single SIS basis function inside the quad. While
    the previous approach certainly aligns better with the non-parametric
    philosophy of this inversion procedure, there may be cases where a
    lack of constraints suggests such an approach.
    The result of this, again an average of 20 individual reconstructions, 
    is shown in Fig.~\ref{fig:smallexpsis}. In this case, the null space
    in the quad region was no longer enabled, but unfortunately only using the 
    overlap fitness did not consistently predict the quad with acceptable
    accuracy. Enabling the RMS fitness measure in conjunction with the standard
    overlap fitness improved the results considerably, and, as the image
    shows, the resulting model can successfully reproduce the quad.

    \begin{figure*}
        \centering
        \includegraphics[width=\WF\textwidth]{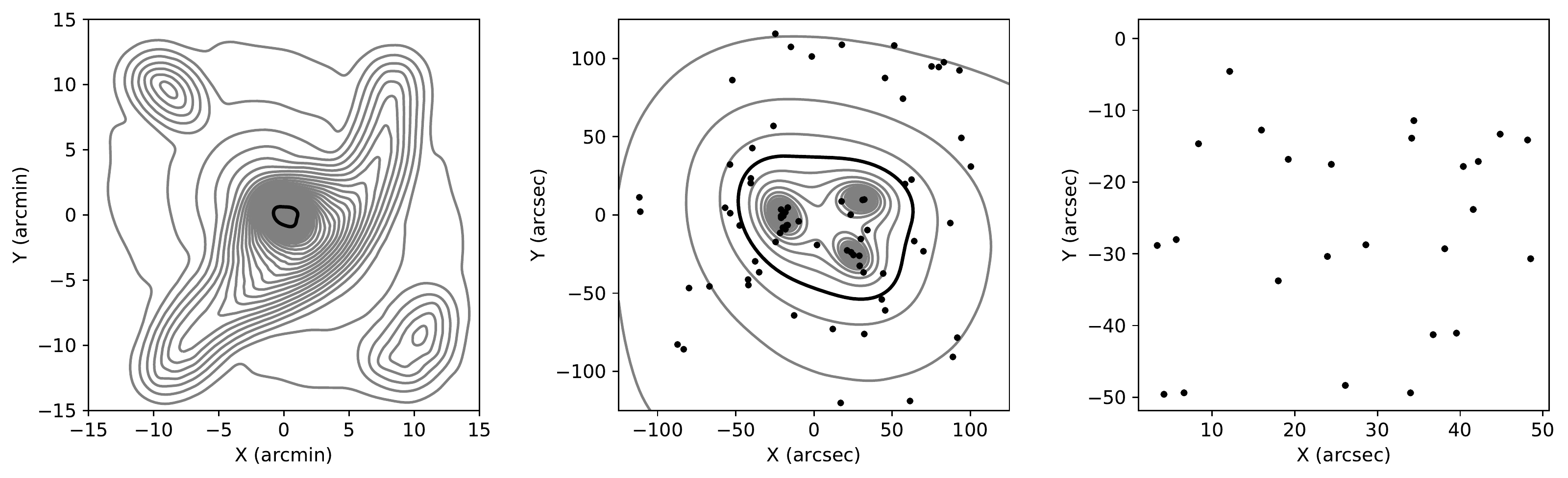}
        \caption{The simulated gravitational lens, at $z=0.4$, used in the weak lensing
        tests. Left panel:~the large scale mass structure that will be probed by ellipticity
        measurements with varying degrees of accuracy. The thick line corresponds to
        $\kappa=1$ for a redshift $z=3$. To make the large scale structure more clearly
        visible, the contours are separated by a small $\Delta\kappa=0.025$.
        Centre panel:~the strong lensing region of the same mass distribution. The thick
        line again corresponds to $\kappa=1$, while the contour spacing is now
        $\Delta\kappa=0.2$. The dots indicate the positions of the 75 point images that
        are included as constraints.
        Right panel:~the 25 point sources that cause the images from the centre panel.}
        \label{fig:weakinput}
    \end{figure*}

    \begin{figure*}
        \centering
        \includegraphics[width=\WF\textwidth]{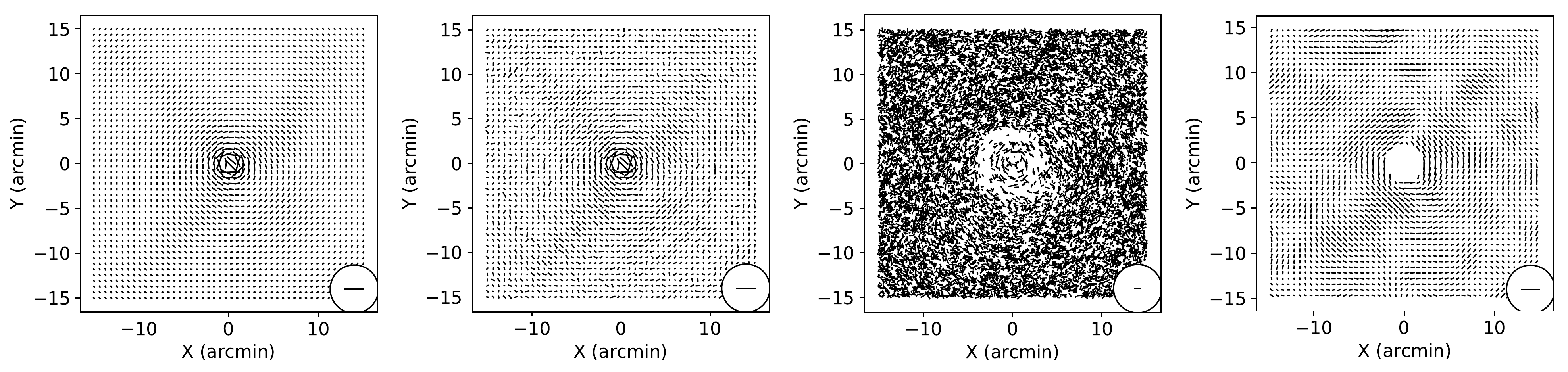}
        \caption{Illustrations of the input ellipticity data used in the three scenarios
        for the weak lensing tests.
        First (left most) panel: in scenario (A), the input is the exact ellipticity
        information calculated from the true model, on a $48\times48$ grid. Here, the
        orientation and sizes for $z=1$ are shown; the inset shows the size of $|\epsilon|=1$.
        Second panel: similar, but for scenario (B), where 25 random source ellipticities
        were transformed at the grid points, and averaged. The result is a more
        noisy version of the previous ellipticities.
        Third panel: for scenario (C), 12,000 source ellipticities at different redshifts
        were transformed. These were binned according to redshift, and on a $48\times48$ grid
        the weighted averages were calculated. 
        The last panel shows the result for the bin corresponding to $z=1.5$, one of the
        inputs in scenario (C). To avoid the less correct weak lensing data affecting the
        more accurate strong lensing constraints, the central $2'$ region was excluded.
        }
        \label{fig:weakinputshear}
    \end{figure*}

    \begin{figure*}
        \centering
        \includegraphics[width=\WF\textwidth]{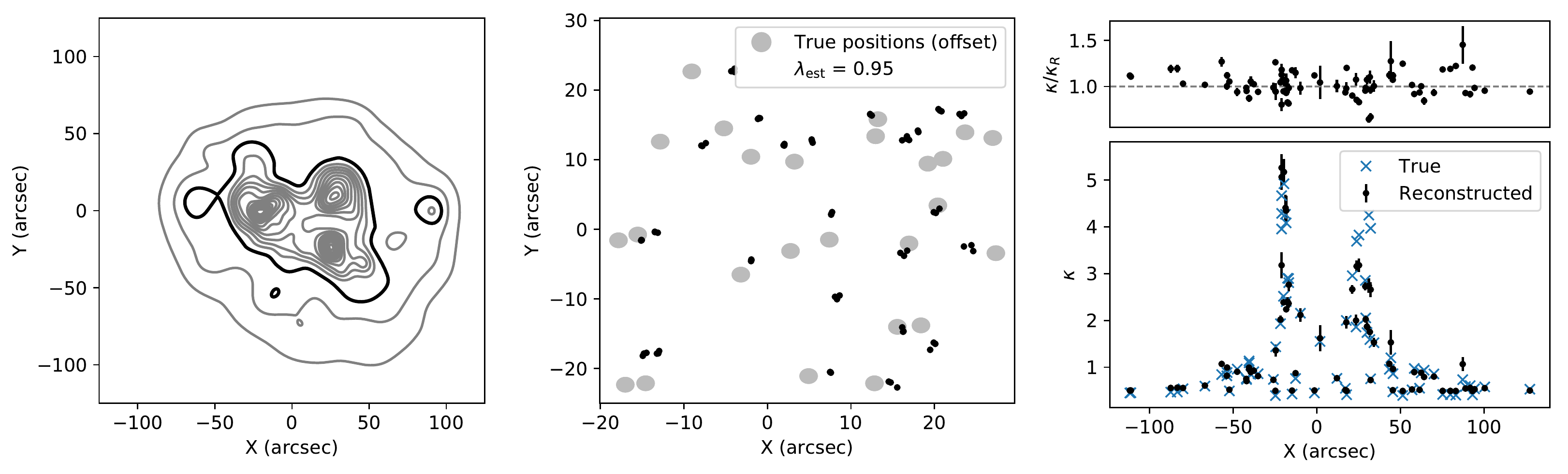}
        \caption{Using only the strong lensing constraints for the simulated
        lens of Fig.~\ref{fig:weakinput}, these results are obtained.
        Left panel:~the average mass map of 20 individual runs,
        where the mass sheet basis function was included. The general features of the
        true mass distribution could be recovered. Centre panel:~comparing the true
        source positions to the back-projected images shows that overall, the scales
        correspond well, although there is slight over-focusing. The diameter of the
        filled gray circles is 2".
        Right panel:~a comparison of the true and
        recovered densities at the image positions, of which the $x$-coordinate is
        used on the horizontal axis, shows no obvious differences in density offset
        or steepness.}
        \label{fig:weakstrongonly}
    \end{figure*}

    \begin{figure*}
        \centering
        \subfigure{\includegraphics[width=\WF\textwidth]{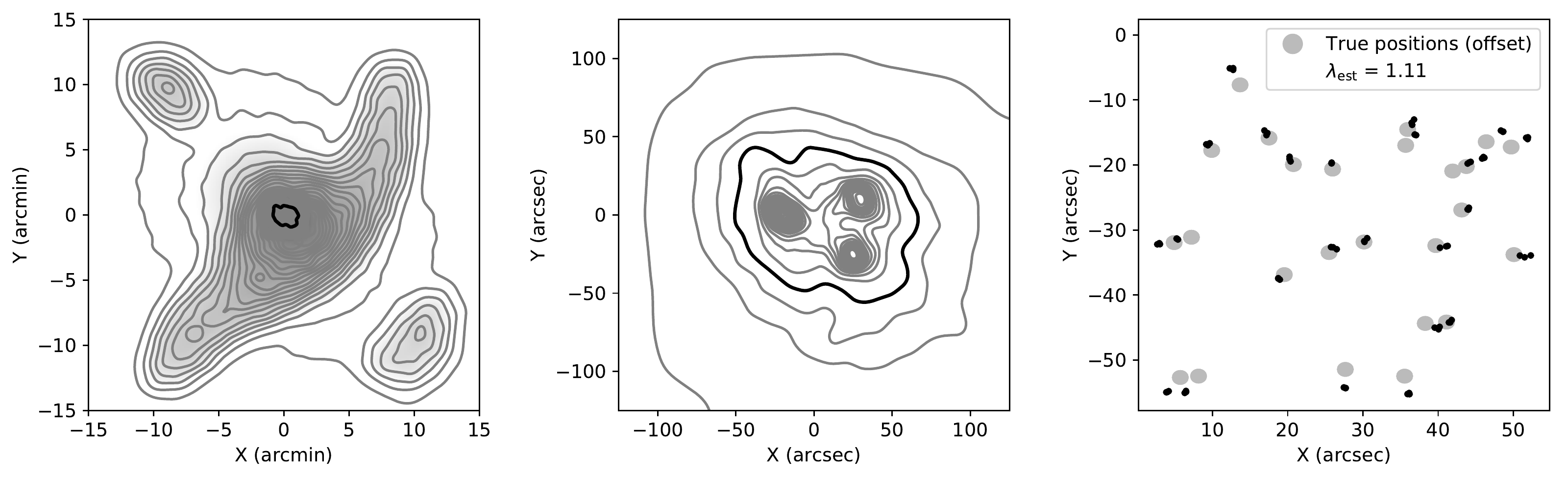}}
        \subfigure{\includegraphics[width=\WF\textwidth]{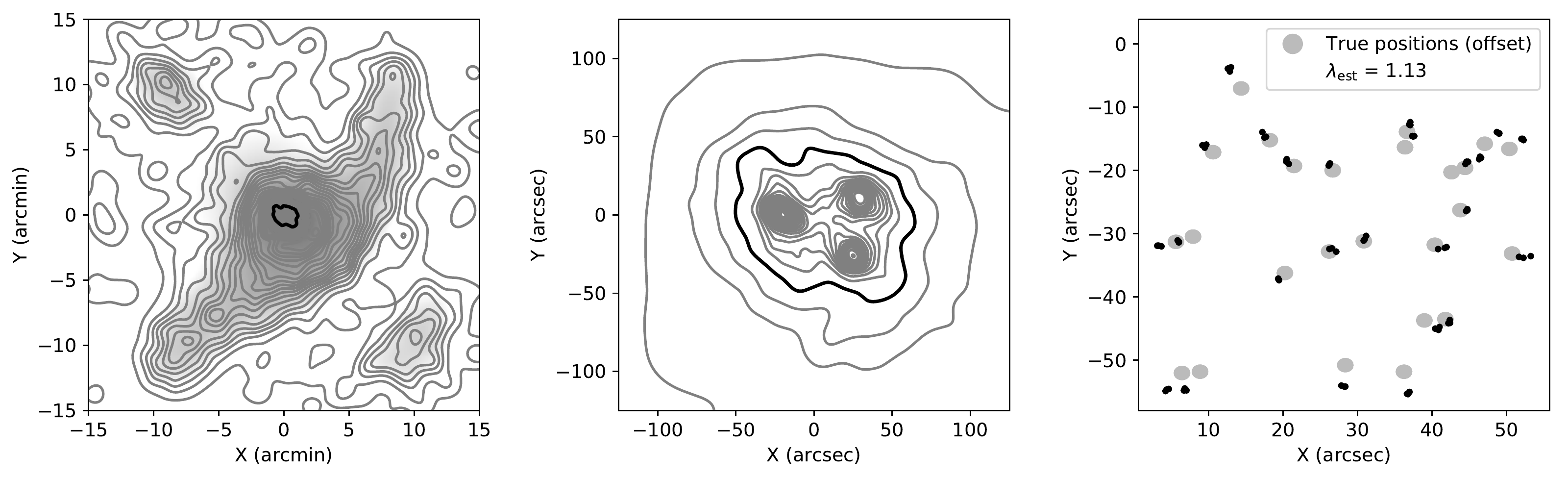}}
        \subfigure{\includegraphics[width=\WF\textwidth]{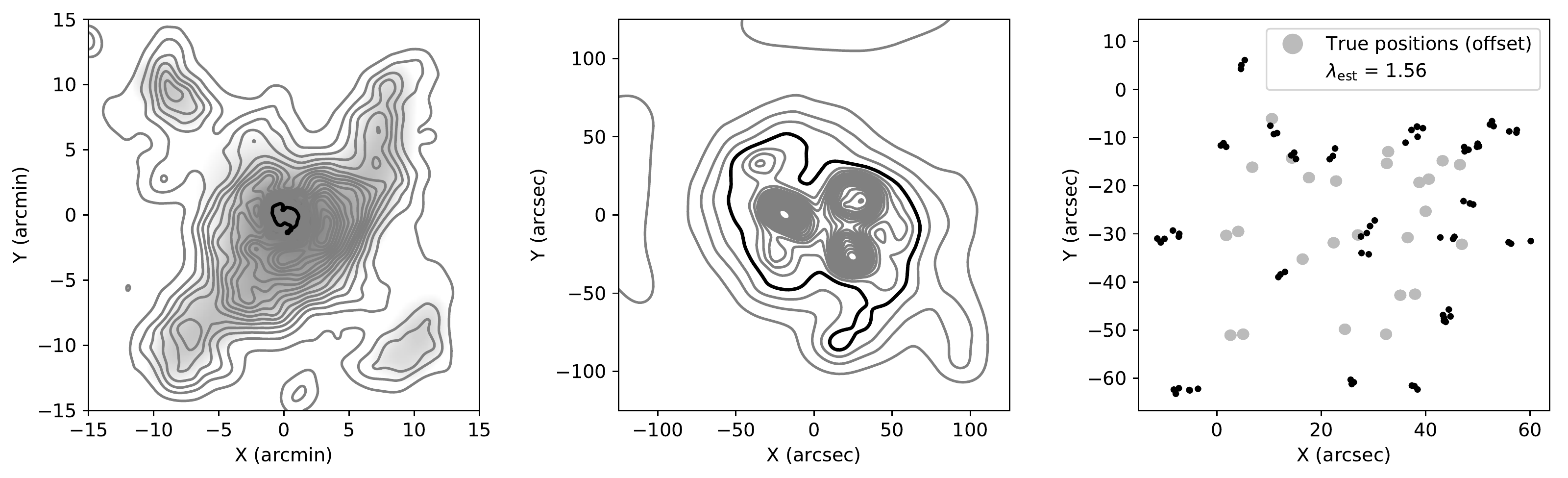}}
        \caption{The three rows correspond to inputs (A), (B) and (C), obtained when the
        mass sheet basis function is not used. The left panels show the wide area reconstruction,
        mainly due to the weak lensing data, where the background shows the shape of the
        true mass density for comparison. To make the small features better visible, a logarithmic gray
        scale was used. 
        The centre panels show the reconstructions of the strong lensing regions, all having
        quite similar features. While the overall shapes are similar, the contour spacing in
        scenario (C) shows a steeper mass map.
        The right panels show the back-projected images in each case. While consistent
        source positions as well as the source plane scales could be recovered for
        scenarios (A) and (B), the scales differing by 11 and 13\% respectively, 
        for scenario (C) neither do the back-projected images overlap
        well, nor is the scale of the back-projected images consistent with the true source
        plane scale. The size of the filled gray circles is again 2".}
        \label{fig:weaknosheet}
    \end{figure*}

    \begin{figure*}
        \centering
        \includegraphics[width=\WF\textwidth]{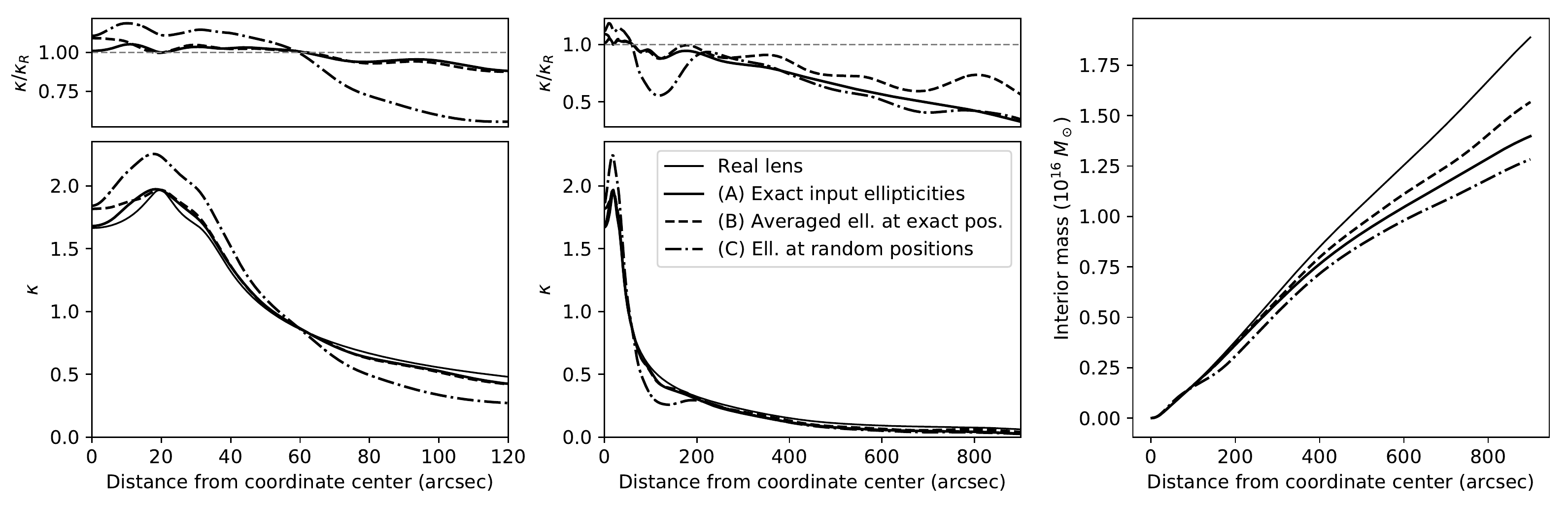}
        \caption{When no mass sheet basis function is used, these profiles are recovered
        in the weak lensing test. Left panel:~the circularly averaged densities in the
        strong lensing region show a good correspondence for inputs (A) and (B), 
        as can also be seen from looking at the relative density $\kappa/\kappa_R$,
        however, the
        steepness and density offset are quite different when the input from scenario (C)
        is used.
        Centre panel:~the circularly averaged densities over the entire weak lensing
        area. Interestingly, even though the strong lensing region is recovered
        incorrectly for scenario (C), beyond $\sim200"$ the profile matches the other
        ones quite well visually, yet upon inspection of the $\kappa/\kappa_R$
        ratio is becomes clear that actually all reconstructions underestimate the real density.
        Right panel:~the integrated mass profiles as well show that in each one of the reconstructions
        the total mass is underestimated.}
        \label{fig:weakprofilesnosheet}
    \end{figure*}

    \begin{figure*}
        \centering
        \includegraphics[width=\WF\textwidth]{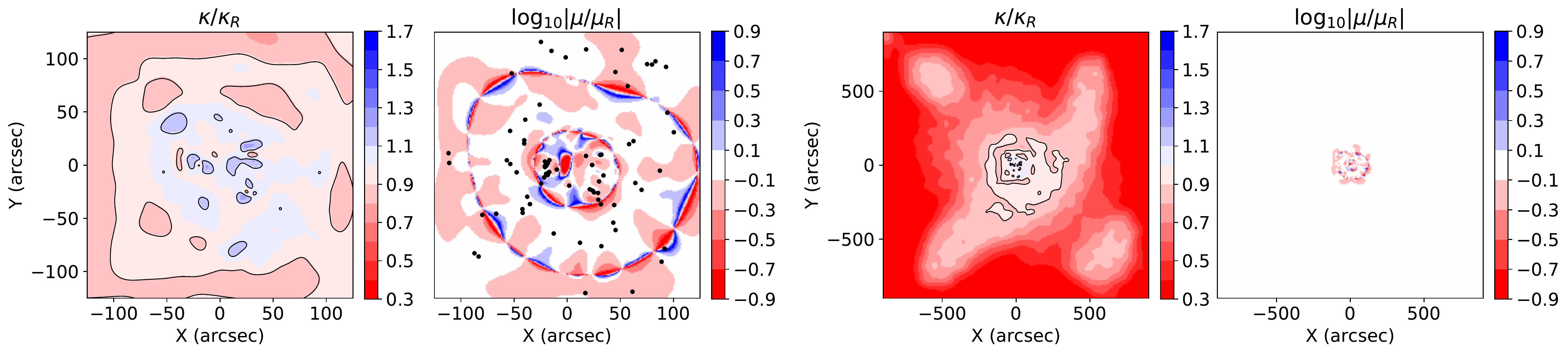}
        \includegraphics[width=\WF\textwidth]{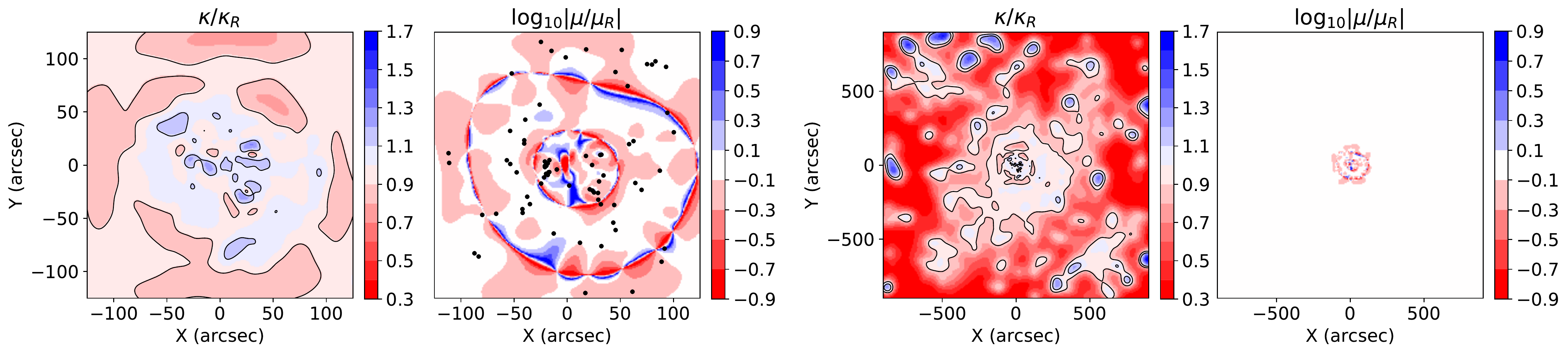}
        \includegraphics[width=\WF\textwidth]{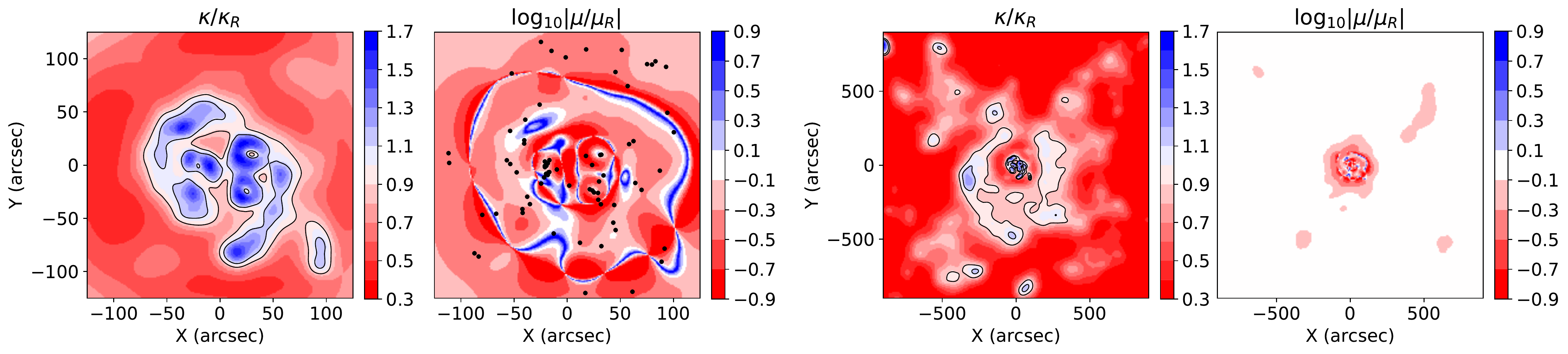}
        \caption{
            The three rows again correspond to scenarios (A), (B) and (C), and, for the
            inversions without enabling the mass sheet basis function, show the fractions
            of recovered densities and magnifications compared to the true ones. The left panels
            show these fractions for the strong lensing region, the right ones for the
            weak lensing region.
            Similar to the circularly averaged profiles, scenarios (A) and (B) show that
            in the strong lensing region the the density traces the true one within
            10\% (black solid lines) in the central part, while the weak lensing maps
            reveal a consistently lower reconstructed density. Not only is this underdensity
            present in (C) as well, but the different steepness in the strong lensing
            region manifests itself as only a very small part in the 10\% range. The
            strong lensing magnifications confirm what could be seen in the plots of the
            source plane scales (right panels of Fig.~\ref{fig:weaknosheet}: matching 
            scales, and therefore magnifications, for (A)
            and (B), while the larger source plane scale for (C) corresponds to a
            lower magnification. For the wider weak lensing regions, the magnifications
            are more similar, in case (C) suggestive of different MSD scale
            factors for both strong and weak lensing regions.
        }
        \label{fig:weakrelativenosheet} 
    \end{figure*}

    \begin{figure*}
        \centering
        \includegraphics[width=\WF\textwidth]{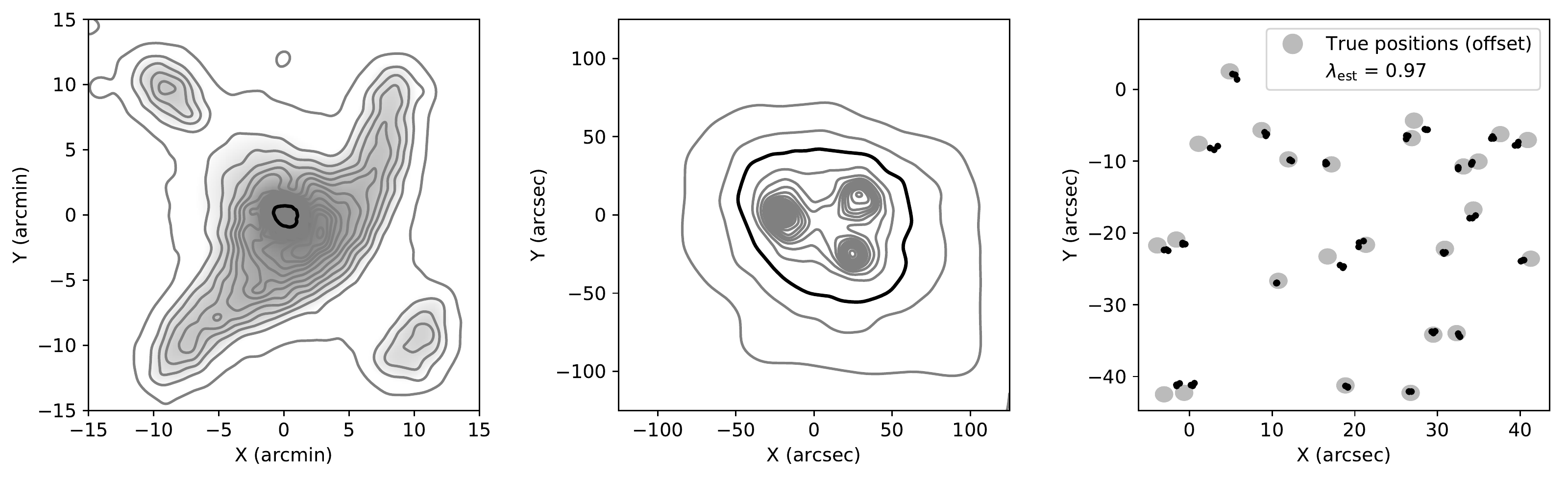}
        \includegraphics[width=\WF\textwidth]{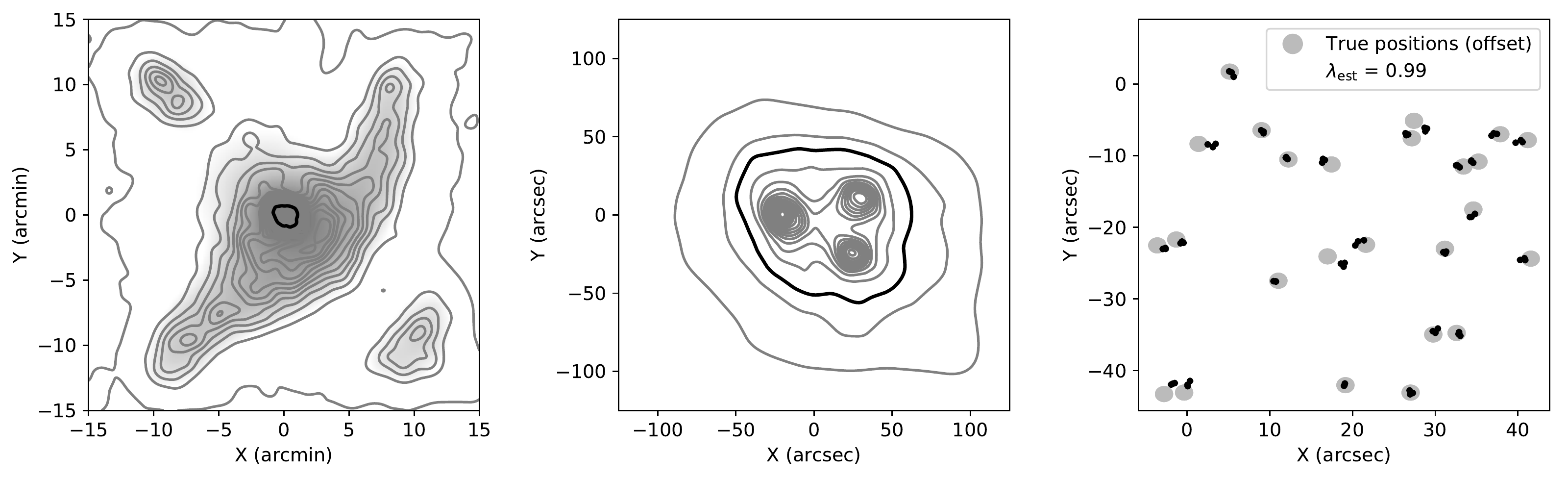}
        \includegraphics[width=\WF\textwidth]{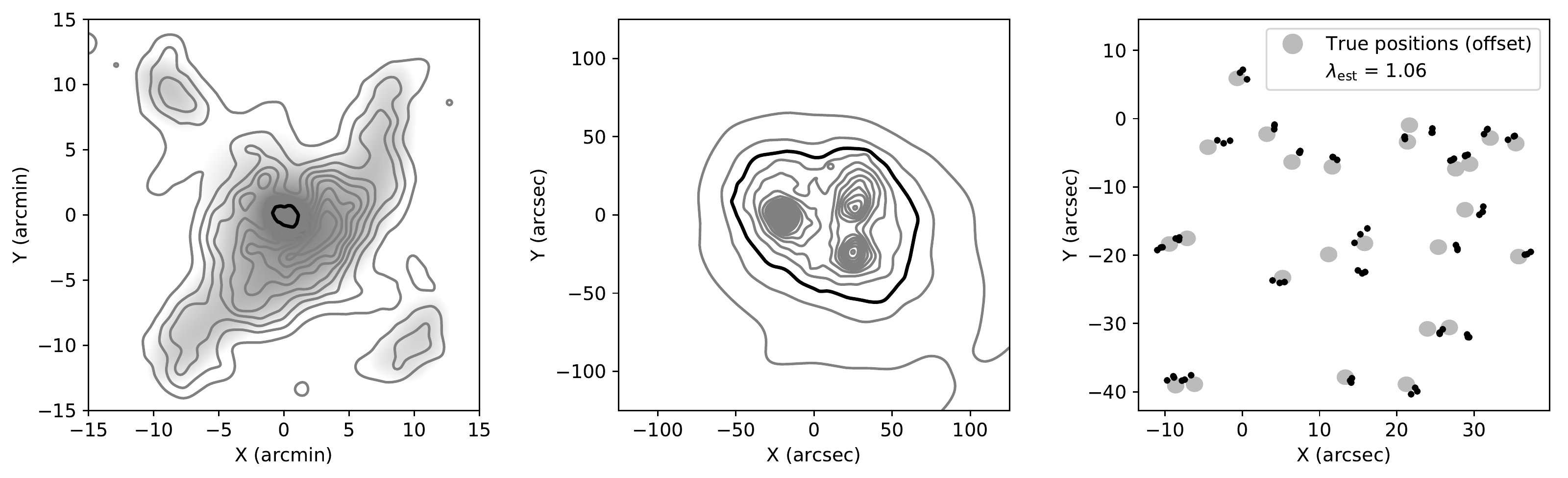}
        \caption{Similar to Fig.~\ref{fig:weaknosheet}, but with the mass sheet basis
        function enabled. In all cases the structure shape in the weak lensing region is
        recovered (left panels), albeit with different levels of accuracy and somewhat 
        different density offsets.
        The recovered strong lensing mass (centre panels) is very similar in each case,
        and the back-projected images overlap well (right panels). The recovered source
        plane scales match the true ones in all three cases, as can be seen
        from their $\lambda_{\rm est}$ estimates, indicating that the MSD scale factor
        is recovered well in the strong lensing region.}
        \label{fig:weaksheet}
    \end{figure*}

    \begin{figure*}
        \centering
        \includegraphics[width=\WF\textwidth]{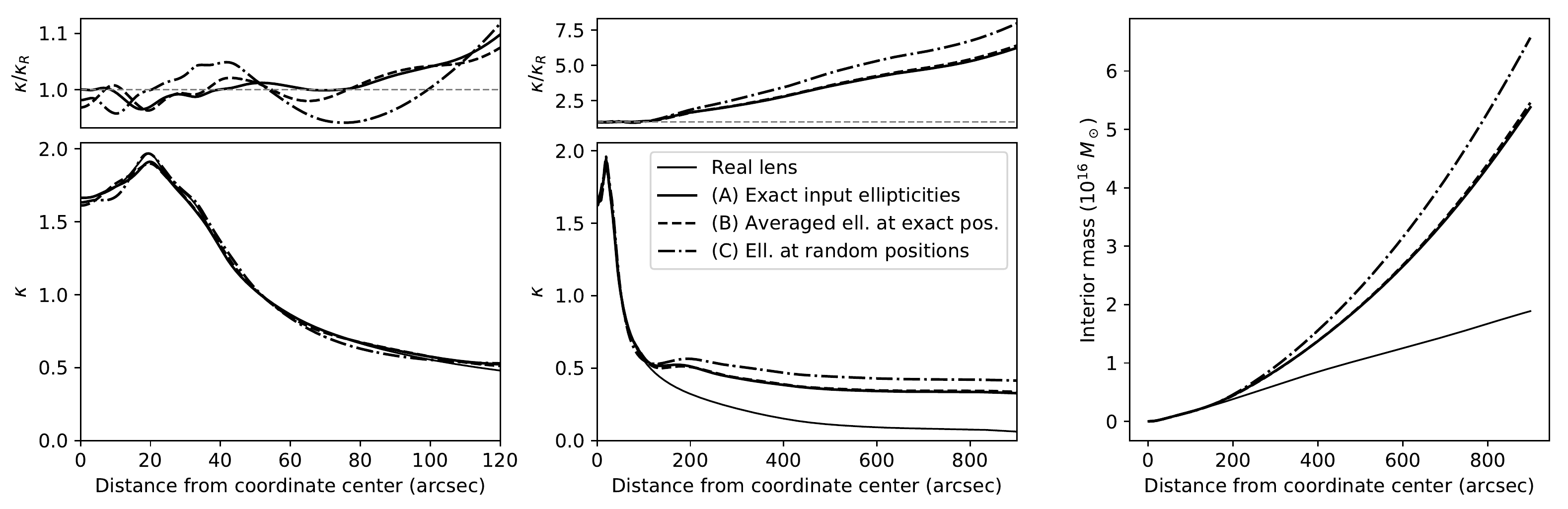}
        \caption{Similar to Fig.~\ref{fig:weakprofilesnosheet}, but with the mass sheet
        basis function enabled. The circularly averaged profiles in the strong lensing
        region (left panel) show a very good correspondence to the true profile in all
        cases, as can also be seen from the plot of the relative density
        $\kappa/\kappa_R$. The profiles for the entire area (centre panel) not only show a considerable
        mass sheet for scenario (C), where the back-projected images show an improved
        reconstruction, but also for (A) and (B). The presence of a considerable mass
        sheet basis function in each case, does not only cause a large ratio of 
        recovered versus true density in the outer regions, but also causes the integrated mass profiles to overestimate
        the correct enclosed masses in all scenarios (right panel).}
        \label{fig:weakprofilessheet}
    \end{figure*}

    \begin{figure*}
        \centering
        \includegraphics[width=\WF\textwidth]{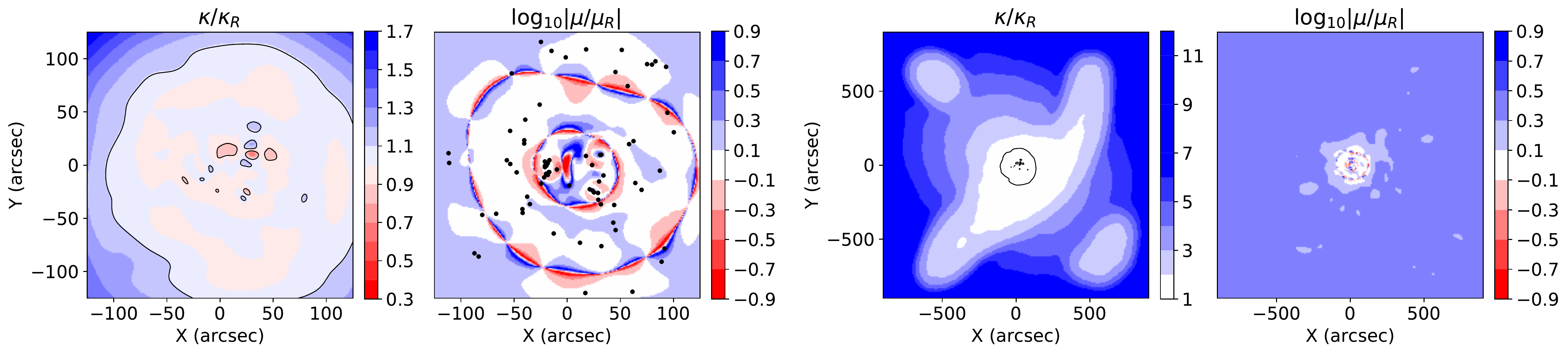}
        \includegraphics[width=\WF\textwidth]{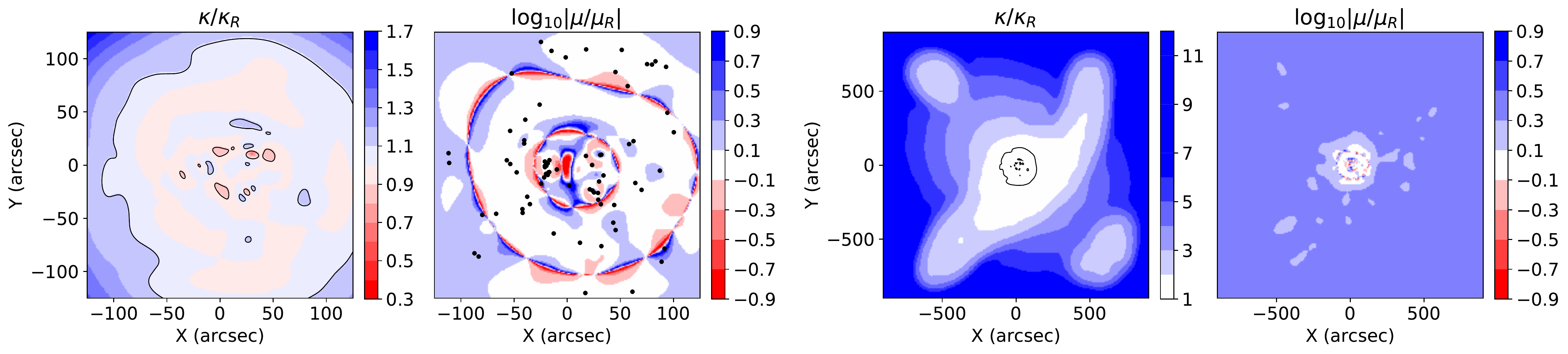}
        \includegraphics[width=\WF\textwidth]{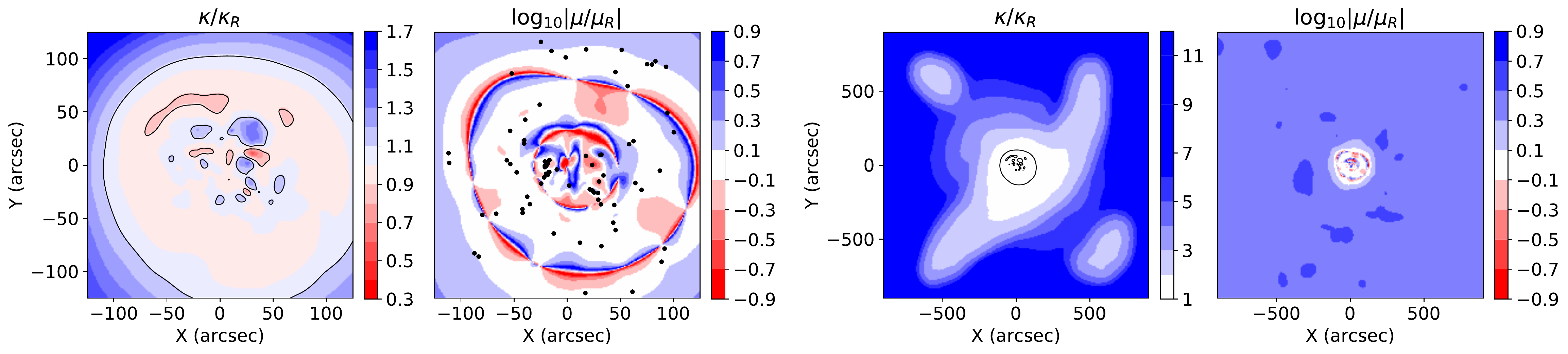}
        \caption{Similar to Fig.~\ref{fig:weakrelativenosheet} but for
        the inversions where the mass sheet basis function was enabled. The strong lensing
        figures indicate a very good correspondence of reconstructed and real mass densities
        over most of the strong lensing region; the matching source plane scales can also
        be seen as matching magnifications. As can be expected from the circularly averaged
        density profiles, the situation is different when considering the weak lensing region.
        Not only is there a considerable relative overdensity in all three scenarios, the
        magnification is also larger than the true one in all three cases, different
        from the one in the strong lensing region.
        }
        \label{fig:weakrelativesheet}
    \end{figure*}

\section{Weak lensing}\label{sec:weak}

    The positions of the multiply imaged sources that can be used to constrain
    the mass density in the strong lensing region, are often available with very
    good accuracy.
    The region in which multiple images are produced is limited however,
    but beyond this, the deformations of background galaxies may still provide
    additional information about the gravitational lens. As the intrinsic orientation
    of the background galaxies cannot be known, this weak lensing signal is
    statistical in nature.

    To work with such information within the inversion framework of \grale, it is 
    assumed that in a pre-processing step, the ellipticity information of the 
    available background galaxies has led to estimates of the average ellipticity
    $\langle\epsilon\rangle^{\rm measured}_i$ at a number of positions $\Vec{\theta}_i$.
    By calculating the reduced shear at these positions, equation (\ref{eq:ellpred})
    allows one to compare these measured values to the ones predicted by the
    model, $\langle\epsilon\rangle^{\rm model}_i$. This suggests the use of the
    fitness measure
    \begin{equation}
        \textrm{fitness}_{\rm WL} = \sum_{i=1}^N w_i 
          \left| \langle\epsilon\rangle^{\rm measured}_i - \langle\epsilon\rangle^{\rm model}_i \right|^2\mcm
        \label{eq:fitnesswl}
    \end{equation}
    where there are assumed to be $N$ such measurements, and weights $w_i$ allow
    control over their relative importances if desired. Alternatively one could
    imagine using all ellipticity measurements directly, preventing overfitting
    the noise by an appropriate choice of a low number of basis functions. While
    the same fitness measure could still be used, this approach will not be 
    explored in this article.
    
    As the reduced shear
    is calculated by dividing the regular shear values $\gamma$ by the factor
    $(1-\kappa)$, these values can become large in regions where $\kappa$ is
    near its critical value, possibly even triggering a division by zero
    error during the optimization. To avoid such regions having a large effect 
    during the course of the optimization, when the GA is still in the process
    of determining the very $\kappa$ map and a near-critical part of an
    otherwise good trial solution could cause it to be discarded, a threshold can be set for $|1-\kappa|$.
    Only points where $|1-\kappa|$ exceeds this threshold are included in the
    summation. In the reconstructions below, this threshold was set to $0.1$.

    \subsection{Simulated lens and reconstructions}

        To study the use of weak lensing data in \grale, the simulated gravitational
        lens shown in Fig.~\ref{fig:weakinput} was used. The shape in the
        strong lensing region (centre panel) is based on a lens model used in
        \citet{Liesenborgs5}, but embedded in the large scale
        structure shown in the left panel. Due to the use of this existing lens model,
        contrary to the other simulations this one used a matter density $\Omega_m = 0.27$.
        The centre panel shows the 75
        point images generated by the sources in the right panel.
        Three scenarios for the weak lensing
        input will be used, each time providing sets of $48\times48$ values for
        $\langle\epsilon\rangle^{\rm measured}_i$, arranged on a uniform grid,
        covering the $30'\times30'$ region.
        
        In scenario (A), the ideal yet unrealistic case,
        these ellipticity values are in fact the exact values 
        calculated from the model using equation (\ref{eq:ellpred}). 
        Furthermore, as weak lensing data at a single redshift do
        not provide enough constraints to fix the MSD scale, not even in
        the case of the exact MSD, let alone a generalization, three different redshifts, $z=1$, $2$
        and $4$, were used to calculate the ellipticities for all grid points.
        The first panel of Fig.~\ref{fig:weakinputshear} shows the orientations
        (see equation (\ref{eq:shearorient})) and sizes of these data points,
        where the length of $|\epsilon|=1$ is shown in the inset.
        For scenario (B), a first degree of randomness was introduced: for each
        grid point, 25 random source ellipticities were transformed using 
        equation (\ref{eq:elltrans}), and these were subsequently averaged.
        As in the previous scenario, this was in fact done for the three
        redshifts of $z=1$, $2$, and $4$. This leads to a more noisy
        version of the true ellipticity field, as can be seen in the second
        panel of the figure. 
        
        In the third scenario, (C), a situation like
        one that could be encountered in practice is considered: similar to 
        the weak lensing information that was made available for the mock 
        clusters Ares and Hera \citep{2017MNRAS.472.3177M}, 12,000 random 
        source ellipticities at random redshifts were transformed by the 
        real gravitational lens model, the result of which is shown in the 
        third panel of the figure. These ellipticities were then distributed 
        over a number of redshift bins, chosen to contain the same interval 
        $0.1$ in $D_{ds}/D_s$ space. 
        For each bin, for each of the grid points, the ellipticities were
        averaged using a gaussian weight function of size 1'; the right panel of
        the figure shows the result for one of the resulting redshift bins.
        The use of a gaussian weight function was also mentioned in
        \citet{1998A&A...335....1L}, to assign more importance to measurements
        close to the grid point under consideration. Other averaging methods
        exist as well, e.g. based on the points that lie inside a grid rectangle
        as in \citet{2006A&A...458..349C}, or even circular regions that are
        different in size, so as to contain a certain number of ellipticity
        measurements \citep{2009A&A...500..681M}. To illustrate the issues
        that may arise from combining strong and weak lensing measurements,
        the chosen method suffices.
        To avoid the lower accuracy weak lensing signal inside the strong
        lensing region interfering with the more accurate strong lensing data,
        a central circular region with a $2'$ radius was excluded in this case.
        As the central region tends to contain bright cluster galaxies, it is furthermore not
        unlikely that only little ellipticity information of background galaxies 
        would be gathered there.

        Note that in all scenarios the ellipticity measurements themselves 
        are still assumed to be entirely correct. The only sources of error
        are the number of measurements available to obtain an estimate of
        the average value (scenario (B)), as well as the spatial distribution 
        of the measured ellipticities (scenario (C)). 
        In settings (A) and (B), no different weights were provided; in (C)
        the number of ellipticity measurements in the redshift bin was used as weight $w_i$.

        To be able to assess the added information provided by the ellipticity data,
        not only in the wider weak lensing region but in the central, strong
        lensing region as well, for reference Fig.~\ref{fig:weakstrongonly} shows the results
        when using only the information from the multiple images, as well
        as the null space. As is usual in these kinds of reconstructions with \grale, the
        aforementioned mass sheet basis function was included. For these
        results, and in the other inversions that follow as well, the average
        solution of 20 runs is shown. As the left panel shows, the general
        features of the central region are recovered, and as the relative
        densities $\kappa/\kappa_R$ are centered on 1, based on the densities
        at the image positions no clear MSD scale factor can be detected.
        As can
        be expected from this correspondence, the scale of the recovered 
        source planes is only slightly different from the true one, as the
        central panel shows, the estimated scale factor $\lambda_{\rm est}$
        differing by only a few percent.

        For the next inversions that we shall consider, apart from the 
        strong lensing fitness measure for point images, $\textrm{fitness}_{\rm WL}$ 
        is enabled in the multi-objective GA as well. The null space fitness 
        measure was not needed to avoid spurious structures in the recovered
        mass maps.
        To provide the Plummer basis functions of which the weights need to be
        optimized, an approach similar to the one used in the example 
        from section~\ref{sec:substruct} is used, but this time for the
        wide area instead of a very small one:
        as in other inversions, the different steps with the subdivision
        grid is used for the strong lensing region, and to be able to model
        the weak lensing region as well, in every step extra basis functions are
        added that cover the weak lensing region, laid out according to a 
        uniform $48\times48$ grid. 
        Originally, a mass sheet basis function was introduced because the strong
        lensing region could contain a non-negligible density offset that
        may be difficult to model using several separate Plummer basis
        functions. Now that the inversion area is much wider, and assuming that
        the density near the border of the region will be low, it makes
        sense to expect the weak lensing signal to recover the overall
        structure, and thereby provide the required density offset inside
        the strong lensing region. The results shown in Figs~\ref{fig:weaknosheet}
        and \ref{fig:weakprofilesnosheet} are therefore the ones obtained
        without enabling a mass sheet basis function.

        The three rows in Fig.~\ref{fig:weaknosheet} correspond to the three
        different scenarios that were described earlier. The top row, in which
        the true average ellipticities were used as constraints -- scenario (A) --
        recovers the wide area structure very well. The centre panel, showing
        the strong lensing area displays a good agreement as well, while the
        back-projected images in the right panel do indicate slightly larger
        source plane scales, although the effect is limited to 11\%. While the input
        ellipticity data is overly optimistic, this scenario does illustrate
        that \grale{} is able to combine weak and strong lensing constraints
        and the code works as expected. The next
        row, where the results of scenario (B) are shown, is actually quite
        similar, but the reconstruction in the weak lensing region is
        clearly more noisy. The third row, showing the results for the more
        realistic scenario (C) is certainly the more interesting one. While
        the overall structure in the weak lensing area is still visible,
        the result does not provide the same accurate representation of the
        shape of the mass as before. This could be expected, as the input
        ellipticity map is diluted by the gaussian smoothing. 
        Looking at the contours in the outer regions, it becomes clear that 
        less mass is recovered there. The strong lensing mass does contain 
        the expected features, but the back-projected images show that not 
        only the reconstruction was not successful in producing well overlapping 
        points, but that the scale set by these points differs by over 50\% from 
        the one set by the true sources.

        This difference suggests the presence of the MSD, which can
        clearly be seen in the left panel of Fig.~\ref{fig:weakprofilesnosheet},
        where this last solution is steeper and has a lower mass offset
        than the true lens, as well as the other two reconstructions. 
        Interestingly, the centre panel indicates that beyond $\sim200"$ 
        the profiles of the reconstructions do resemble the true profile, 
        although the relative density $\kappa/\kappa_R$ of
        recovered to real models shows that it is consistently and
        increasingly underestimated. This can also be seen in 
        the right panel, showing the total enclosed mass
        vs. radius, indicating that not all mass has been captured by the
        reconstructions. This is not only the case for the reconstruction
        using input (C), but even for (B), as well as for the very correct
        looking solution for (A).

        Fig.~\ref{fig:weakrelativenosheet} shows
        relative densities and magnifications of recovered and real lens models,
        in both strong and weak lensing regions. In scenarios (A) and (B)
        the main difference is the density in the wider, weak lensing area,
        but for (C), the strong lensing region as well shows considerable
        differences. The difference in source plane scale can be seen as
        a consistently lower magnification. While the difference in steepness
        and mass offset in the strong lensing region, as well as the different
        magnification can all be attributed to the MSD, the fact that the 
        magnification deviates with a different factor in strong and weak
        lensing regions suggests a different MSD-like effect in both regimes.

        Disabling the mass sheet basis function was prompted by
        the idea that the reconstruction in the weak lensing area would provide
        any mass sheet like effect that might otherwise be difficult to account
        for. But as the previous reconstruction for input (C) still seems to
        suffer from the MSD, and this was not the case for the strong lensing only
        reconstruction, it is interesting to see the effect of the
        inclusion of such a basis function. Similar to the previous figures,
        Figs~\ref{fig:weaksheet}, \ref{fig:weakprofilessheet} and \ref{fig:weakrelativesheet} show the
        results for these inversions, this time with a mass sheet basis function
        enabled. The left panels of Fig.~\ref{fig:weaksheet} show that in all
        three cases the shape of the wide area mass distribution does seem to be
        recovered, albeit with different mass offsets, slopes and level
        of detail. The centre panels show strong lensing masses that are very
        similar to the true mass distribution there, and the right panels 
        indicate that the recovered source plane scales correspond very well to the
        true ones.

        That the MSD scale factor is obtained correctly, at least in the strong lensing
        region, can also be seen in the left panel of Fig.~\ref{fig:weakprofilessheet},
        where all profiles now have similar steepness and offset,
        and in the left part of Fig.~\ref{fig:weakrelativesheet}.
        In the right part of that figure, showing the wider area, 
        as well as in the centre panel of Fig.~\ref{fig:weakprofilessheet}, the
        presence of the mass sheet basis function becomes very obvious however.
        The integrated mass, in the right panel, is therefore clearly larger
        than the correct one in all three cases. Whether or not there is
        non-negligible density in the outer regions, and using a mass sheet
        component may be warranted, will depend on the situation. As this example
        illustrates, it may however not be straightforward to determine this
        automatically.
        While the correctly estimated source plane scales in the strong
        lensing region can be seen
        as matching magnification ratios in the left part of Fig.~\ref{fig:weakrelativesheet},
        the situation is clearly different in the weak lensing region. The
        fact that there the magnification is consistently larger than the true one
        is again suggestive of different MSD-like scale factors in strong and weak
        lensing areas.

        \begin{figure*}
            \centering
            \includegraphics[width=\WF\textwidth]{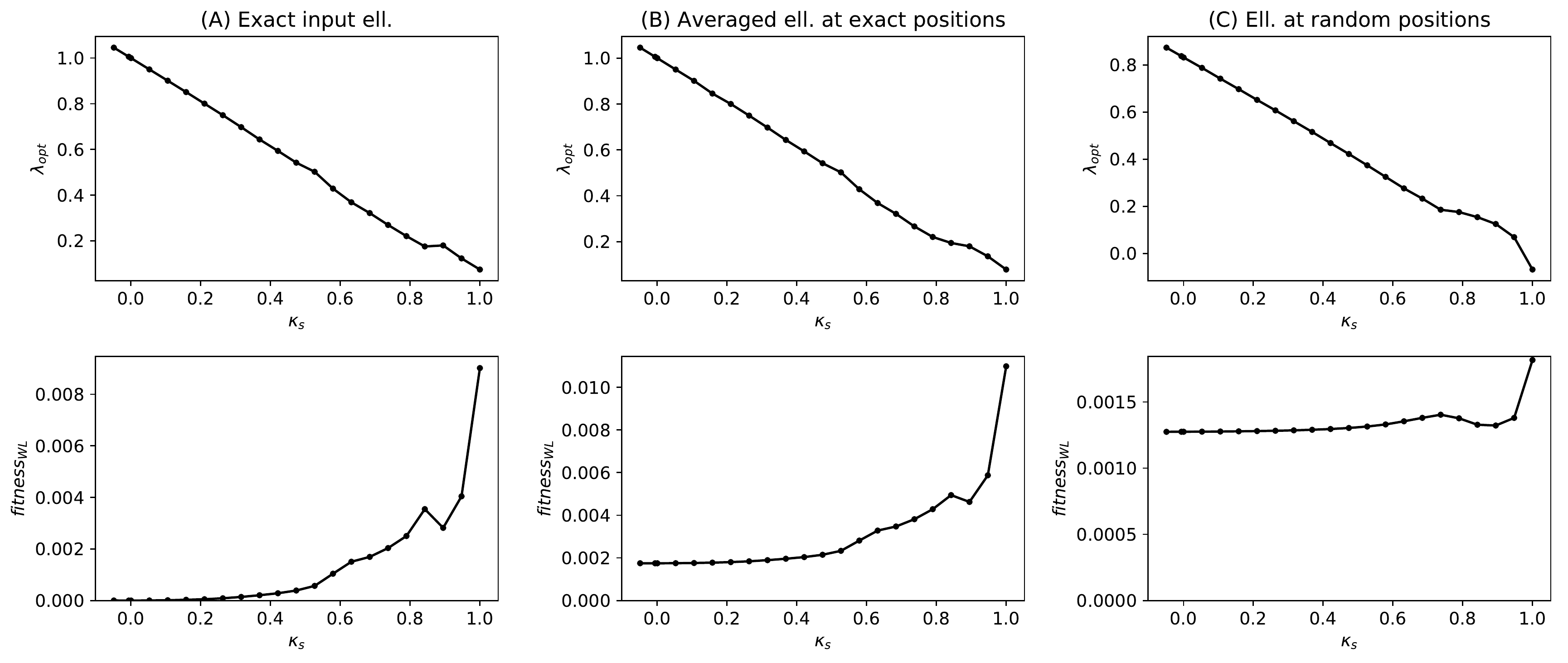}
            \caption{Based on the lens model with density $\kappa_0$ from
            Fig.~\ref{fig:weakinput}, a new model is constructed: starting
            from a mass sheet with specific density $\kappa_s$, a factor
            $\lambda_{\rm opt}$ is determined numerically so that the new
            model $\kappa_1 = \lambda_{\rm opt}\kappa_0 + \kappa_s$ yields 
            the lowest weak lensing fitness value. The columns indicate the
            different input scenarios for which this test was performed, the
            top row shows how, similar to the exact MSD (equation (\ref{eq:msd})),
            there's a nearly linear relationship between $\kappa_s$ and
            $\lambda_{\rm opt}$, while the bottom row indicates that depending
            on the precise scenario, the relative fitness values do not change
            much.
            }
            \label{fig:weakmsd}
        \end{figure*}

    \subsection{Mass-sheet-like degeneracy}

        The fact that with this mass sheet basis function enabled, the shape
        of the mass density in the weak lensing area is still recovered well,
        is of course in itself a manifestation of the MSD. Whereas the last
        images show that in the strong lensing region the MSD scale
        factor is obtained correctly, in the weak lensing region a different
        one is obtained.

        To better understand why the weak lensing data, containing information
        about different redshifts in each scenario, is not able to constrain 
        this contribution better, a simple experiment is performed: starting from the
        true lens model $\kappa_0$ from Fig.~\ref{fig:weakinput}, a new one is constructed
        in the following way. First, a mass sheet with a specific fixed density
        $\kappa_s$ is chosen. Next, a scaled version of the true lens is added
        where the scale is determined numerically to be the one that minimizes
        $\textrm{fitness}_{\rm WL}$, leading to the lens with the mass density
        \begin{equation}
            \kappa_1(\Vec{\theta}) = \lambda_{\rm opt} \kappa_0(\Vec{\theta}) + \kappa_s \mpt
        \end{equation}
        The results are shown in Fig.~\ref{fig:weakmsd},
        where columns refer to the different input scenarios (A), (B) and (C),
        the top row shows the scale factor $\lambda_{\rm opt}$ for each mass
        sheet density $\kappa_s$, and the bottom row shows the effect on the
        fitness measure. Note that these tests calculate $\textrm{fitness}_{\rm WL}$
        for various lens models, constructed in the way described here, 
        but do not use the GA based inversion procedure.

        Similar to \citet{1997A&A...318..687S}, where a distribution of redshifts
        is shown to still allow an MSD, the top row also shows a nearly linear 
        relation for the best scale factor with respect to the density of the 
        mass sheet. 
        For scenarios (A) and (B), the lowest fitness value is for the true
        density map, or $\kappa_s = 0$, indicating that in those cases the
        different redshifts successfully break the MSD. Note however that
        in scenario (C), while the lowest fitness value in the experiment
        is still for $\kappa_s = 0$, the value for $\lambda_{\rm opt}$ 
        actually differs from unity, i.e. a scaled mass map has a better fitness
        value than the true $\kappa_0$. Note that scenario (C) was obtained
        by using a gaussian weight function; in \citet{1998A&A...335....1L}
        it is shown that the smoothed shear field then actually corresponds
        to a density map that is convolved with the same weight function,
        which can explain this observation.
        As the noise increases, the relative effect 
        on the fitness becomes increasingly less prominent, and furthermore, 
        the effect studied here is when the exact lens $\kappa_0$ is used, but 
        in practice this is not the density that is available during the 
        optimization. Instead, multiple basis functions are used to optimize the 
        fitness measure, which may make the sensitivity to $\kappa_s$ even worse. 

        As also noted by \citet{2004A&A...424...13B}, this experiment indicates 
        that while in principle weak lensing data for multiple redshifts can break the 
        MSD, in practice it is much less evident. It is therefore not surprising that
        when combining this information with the strong lensing data, and allowing
        for a mass sheet basis function, it is in fact the strong lensing data
        that dictates its value, causing a scaled mass density in the larger,
        weak lensing region.

        For another way to look at this, let us assume that the
        strong lensing constraints and weak lensing constraints do not overlap.
        Based on the available data, one could perform a reconstruction for the
        strong lensing region, corresponding to a projected potential $\psi_{\rm SL}$.
        Similarly, one could perform a reconstruction based solely on the weak
        lensing measurements, leading to $\psi_{\rm WL}$. The former
        need not be valid in the weak lensing region, and the latter will
        only provide a very rough estimate in the strong lensing region. Let us
        next consider only the relevant parts of these potentials, $\psi_{\rm SL,central}$
        for the strong lensing region, and $\psi_{\rm WL,nocentral}$ for the weak
        lensing region, both covering their respective constraints. As any value
        can be added to the potentials without affecting observables, one can
        imagine creating a $\psi_{\rm SL+WL}$ which is in essence $\psi_{\rm SL,central}$
        in the central region, and $\psi_{\rm WL,nocentral}$ further out, using some
        interpolation in between. The combined potential will perform as well
        as both separate potentials in their respective regions, but the way the
        interpolation has occurred, obviously has an effect on the resulting mass
        density in between these regions, possibly allowing multiple equivalent
        lenses.
        Additionally, one can imagine that, if the MSD is not fully broken in
        neither strong nor weak lensing regions, both lensing potential parts
        can be first modified to create equivalent ones, and only then combined.
        The MSD scale factor would not even need to be the same for both parts,
        leading to even more equivalent lens models.

        It is also interesting to note that the weak lensing constraints on the
        lensing potential are purely local ones, relating to only its curvature.
        One could therefore imagine different MSD contributions for different
        regions, for example a different mass sheet contribution as one moves
        further from the centre. As different regions may be differently sensitive
        to the MSD, this may be an additional effect to take into account, also
        easily affecting the total mass estimate. 

    \subsection{Ring-like structure}

        The circularly averaged density plots from Figs~\ref{fig:weakprofilesnosheet}
        and ~\ref{fig:weakprofilessheet} are not centred on one of the three
        mass peaks, but on the centre of the coordinate system. This highlights
        a feature which might otherwise be lost in the averaging procedure: in
        Fig.~\ref{fig:weakprofilesnosheet}, the reconstruction for input (C)
        clearly shows a relative overdensity around $200"$. For scenarios (A)
        and (B), the effect is much less pronounced, but the reconstructed
        profile still deviates from the true one on the boundary between strong
        and weak lensing regions. In Fig.~\ref{fig:weakprofilessheet}
        a similar effect can in fact be seen in all scenarios.

        Such effects have been encountered in other works where strong
        and weak lensing measurements were combined. In 
        \citet{2007MNRAS.375..958D} these ring-like structures were
        also perceived, followed by the argument in
        \citet{2011A&A...535A.119P} that such effects can be caused
        by overfitting. Interestingly, the work of \citet{2007ApJ...661..728J}
        identified a dark matter ring in Cl~0024+17 based on a combination
        of strong and weak lensing data. While the authors make a plausible
        argument for the possible origin of such a structure, further
        observations to confirm the existence are still awaited.

        The discussion above about combining weak and strong lensing
        regions, each possibly having its own MSD, provides a possible 
        explanation for such ring-like features,
        perhaps not surprisingly located on the border between the two regimes. 
        More importantly, this has no intrinsic relation to the amount
        of overfitting, although depending on the precise method used,
        the degree of overfitting may affect the MSD in both regions,
        possibly separately, and can therefore be seemingly related to
        the introduction of such a feature. 

\section{Discussion and conclusion}\label{sec:discussion}

    In this article we have reiterated the current capabilities of the
    inversion procedure in \grale, a genetic algorithm based method, and 
    investigated a number of upgrades that are now available. In this
    method, a number of criteria can be optimized concurrently,
    a common combination being the overlap of back-projected images, as well
    as the absence of unobserved, extra images, i.e. the null space. While
    the combined optimization of multiple aspects is certainly reminiscent of
    a regularization procedure, the criteria are not optimized together in the
    typical fashion of regularization, i.e. by combining multiple goodness-of-fit 
    measures into a single number, with suitably chosen scale factors for the
    different aspects. Instead, a multi-objective genetic algorithm looks for
    a common optimum of the different so called fitness measures.

    Which of the available fitness measures should be used will depend on the
    specific gravitational lensing system that is being studied. 
    The basic requirement is having multiply
    imaged sources, but the addition of null space information may or may not cause
    improvements, depending on the amount of images, their spatial distribution
    and their redshifts. Because the 
    inclusion of null space information causes the computational complexity to
    increase considerably, it is certainly worth exploring whether disabling the 
    null space fitness already yields acceptable solutions, 
    as more and more multiple image systems tend to become the norm:
    typically the null space will be most useful for the
    systems with fewest sources. Similarly, the fitness values describing a critical 
    line penalty from \citet{Liesenborgs4} and \citet{Liesenborgs5} were so far only 
    needed in these works, where only very few sources were available.

    Because of the flexibility of the basis functions used, in principle this now allows
    an approach that resembles the working of simply parameterized inversion methods:
    place basis functions such as a pseudo isothermal elliptical mass distribution 
    (PIEMD, e.g \citet{2007arXiv0710.5636E}) based on the observed cluster members
    and optimize their contributions using the GA. Note however that, while this approach
    is certainly feasible, the GA can only change the weights of these basis
    functions, and not other parameters, such as ellipticity or orientation.
    How useful this feature turns out to be, for example to combine models of
    visible galaxies with a more non-parametric contribution, still needs to
    be investigated.

    Time delays can provide very useful information about the overall
    lensing potential, and the time delay fitness measure proposed in
    \citet{Liesenborgs5} was evaluated more thoroughly. The test was similar to
    the one used in \citet{2012MNRAS.425.1772L,2019A&A...621A..91W}, where a lens with an
    elliptical version of a Navarro-Frenk-White (NFW; \citet{1996ApJ...462..563N})
    mass distribution was used to gauge the effect of the addition of
    time delay information to the images of only a single source. Here, it was
    found that when adding time delay information to the images of all sources, the
    old fitness measure can over-focus the images whereas a newly
    proposed expression appears to avoid this. The same effect, although
    to different degrees, was seen in other tests as well, both
    with point images and extended images: the new
    time delay fitness measure yielded time delay predictions that were
    more compatible with the input time delays, and the resulting mass
    distributions suffered less from MSD effects. 
    It is interesting to note that if the images of only one source were
    equipped with time delays, the over-focusing effect did not appear
    for the A1689-based test.

    Especially in the application to MACS~J1149.6+2223 \citep{2019MNRAS.482.5666W}
    it was clear that the subdivision grid based procedure cannot always
    provide adequate resolution to account for all observed features. The
    updates to the inversion code include more flexibility regarding the basis
    functions used, thereby allowing small scale structures to be introduced
    when studying a gravitational lens simulation with similar features
    as MACS~J1149, in turn causing the
    observed images to be predicted more accurately. In this approach, one
    can again use a number of basis functions to be able to describe a more
    general small scale mass distribution, or use a single profile, e.g. a SIS,
    guided by the observed light. As such small scale substructures may not have
    many constraints, it will depend on the specific case
    at hand which approach may be more appropriate. Note that in both cases,
    a manual intervention to account for small scale features is required,
    making the method less automatic, and perhaps somewhat less free-form.
    Investigating ways to automate this -- e.g. adding substructure due to
    an underestimate of image multiplicity -- will be the topic of further
    research.

    While it appeared relatively straightforward to add weak lensing 
    constraints as an extra fitness measure, the experiment shown revealed
    some interesting aspects. In Fig.~\ref{fig:weaknosheet}, the first two scenarios 
    show that if the quality of the ellipticity measurements supplied to the 
    inversion is good, even allowing for some noise, the weak and strong 
    lensing data can be combined quite successfully. However, even in the
    case where the ellipticity measurements are exact, the border between
    both regimes is visible in Fig.~\ref{fig:weakprofilesnosheet}. An
    incompletely recovered total mass further indicates that the MSD 
    scale factor was not fully retrieved:
    while the true model has a lowest convergence value of
    $\kappa = 0.027$, the reconstructions have a minimum that is an order
    of magnitude smaller. As the weak lensing signal is further diluted in
    scenario (C), these
    measurements no longer provide a strong enough constraint in the
    strong lensing region to obtain an accurate MSD scale there. The result for the strong
    lensing region is then what is typically seen if the mass sheet basis
    function is not included: the algorithm does not succeed well in
    finding a model that creates overlapping back-projected images, as
    it is not straightforward to create an overall mass sheet like effect
    using several different Plummer basis functions. In scenarios (A)
    and (B) the weak lensing constraints on the other hand were effective
    in creating the right environment for the strong lensing data.
    In this sense, enabling the use of the mass sheet basis function can
    help in providing the right strong lensing environment, in all three
    scenarios. In that case, one can still learn about the general shape
    in the weak lensing region, but the precise mass density is
    clearly overestimated. If one is only interested in the strong lensing
    region, the inclusion of weak lensing data can be seen as having a sort
    of stabilizing effect, not unlike what was mentioned in \citet{2007MNRAS.375..958D}.

    Breaking the MSD is not straightforward, not necessarily in the strong
    lensing region, not in the weak lensing region, and not in the combination
    of the two. Ultimately, the precise inversion method used, and in particular
    what kind of prior information it uses about the mass density that is
    expected, can make an important mark on the way the degeneracy is
    affected. This is what was visible in the examples shown, in various
    ways. In scenario (C) without the mass sheet basis function, the MSD
    scale factor was not obtained
    correctly in the strong lensing region, but was 
    nevertheless combined with a more correct looking weak lensing 
    reconstruction -- although still a small effect leading to a mass
    deficit remained. In scenarios (A) and (B), with the inclusion
    of the mass sheet basis function, the strong lensing constraints
    favored the algorithm to proceed towards a solution where a considerable
    mass sheet was present, even though a good solution was shown to
    be available without this basis function. The presence of such a mass
    sheet then obviously affected the weak lensing mass density as well.

    Regarding the addition of weak lensing data to the strong lensing
    oriented \grale, the results are somewhat mixed at this point. If the
    quality of the ellipticity information is good, a reasonable weak \& strong
    inversion at least seems possible, even though possible artifacts
    at the boundary between strong and weak lensing regions should always
    be kept in mind, as well as the difficulty in breaking the MSD globally. 
    If one is only interested in the strong lensing region, weak lensing
    data, even of less quality, can help constrain the environment, leading to
    good results in said strong lensing region. Furthermore, while the
    inclusion of weak lensing data incurs additional computational
    complexity, it does appear that they help constrain the mass density
    in such a way that including the null space, and its associated
    computations, can be avoided. To obtain an inversion result that
    appears more plausible over the entire region, in the absence of
    better quality weak lensing data, some further experimentation is
    needed. Interesting avenues include using the weak lensing only
    reconstruction as a base lens, and let the inversion procedure
    look for corrections to this model, either with all positive Plummer
    basis functions, or allowing negative ones as well. Related to this method,
    but allowing the contribution of this base model to vary, said
    model in its entirety could be added as one of the basis functions
    in the optimization. A different approach would be to tweak the GA
    itself, influencing the way the parameter space is explored; possibly
    by laying emphasis on the weak lensing reconstruction early on, and 
    later shift to the strong lensing constraints to capture the details
    in the central region.    
    The updates to the framework have introduced
    much flexibility, but thereby also multiple approaches to
    handle these, and other, issues. Exploring these further is the
    topic of future investigations.

\section*{Acknowledgments}
    
    We gratefully acknowledge the \lenstool{} team for making their models 
    publicly available on-line.
    JL acknowledges the use of the computational resources and
    services provided by the VSC (Flemish Supercomputer Centre), funded by the
    Research Foundation - Flanders (FWO) and the Flemish Government -- department
    EWI.
    JW gratefully acknowledges the support by the Deutsche Forschungsgemeinschaft 
    (DFG) WA3547/1-3.
    SDR acknowledges financial support from the European Union's Horizon 2020 
    research and innovation programme under the Marie Sk{\l}odowska-Curie grant 
    agreement No 721463 to the SUNDIAL ITN network.

\appendix
\section{Source position offsets}\label{app:srcpos}
In case a source plane is rescaled by means of the MSD,
see eq (\ref{eq:msd}), the source position
in general will change as well, so to compare scales one would
have to take this change in position into account. Using a mass
sheet, the rescaling will be centred on the origin of the coordinate
system, but in case a mass disc is used, which has a similar effect
as long as all images are covered, the centre of the disc determines
the centre of the rescaling operation. This way, for a lens that
differs by the MSD, many different source positions can be seen to
correspond to the same image positions.

To understand that the source position can also change
when there is no apparent MSD involved, one merely has to use two
such MSD constructions: one that uses a scale factor $\lambda$,
followed by one with a scale factor $\lambda^{-1}$. Using a mass
disc for each, but with a different centre, allows one to obtain
a different source position that corresponds to the same images.
The net effect of the procedure is a mass density that is exactly
the same in a central region, which can be made arbitrarily large,
surrounded by a ring-like structure. While the source position
has changed, none of the observables have.

That the source plane offset has no direct meaning can also be seen by
noticing that both a given lensing potential $\psi_0(\Vec{\theta})$ and
\begin{equation}
    \psi_1(\Vec{\theta}) = \psi_0(\Vec{\theta}) + \Vec{a}\cdot\Vec{\theta}
\end{equation}
correspond to the same images and mass density, but with a shift in 
source plane \citep{1998A&A...337..325S}. This can be seen to be a special
case of the more general equation (21) in \citet{2018A&A...620A..86W} which
shows the change in potential that corresponds to a shift in source
plane position.

\bibliographystyle{mnras}
%\bibliography{paper}

\bsp 
\label{lastpage}

\end{document}